\documentclass[12pt]{article}

\usepackage{amssymb}
\usepackage{amsmath}
\usepackage{amsbsy} 

\usepackage{graphics}

\begin{document}

\title{\bf OPE coefficient functions in terms of composite
operators only. Nonsinglet case}

\author{A.V. Kisselev\thanks{E-mail: alexandre.kisselev@mail.ihep.ru} \
and V.A. Petrov\thanks{E-mail: vladimir.petrov@mail.ihep.ru} \\
\small Institute for High Energy Physics, 142281 Protvino, Russia}

\date{}

\maketitle

\thispagestyle{empty}

\bigskip

\begin{abstract}
A new method for calculating the coefficient functions of the
operator product expansion is proposed which does not depend
explicitly on elementary fields. Coefficient functions are defined
entirely in terms of composite operators. The method is
illustrated in the case of QCD nonsinglet operators.
\end{abstract}

\bigskip
\centerline{PACS numbers:  11.10.-z, 11.10.Gh, 12.38.-t}

\clearpage

\section{Introduction}

The structure functions of deep inelastic lepton-proton scattering
(DIS) are related with an absorptive part of the matrix elements
\begin{equation}\label{002}
i \int \! d^4x \,\mbox{e}^{iqx} \langle \, p \, |\mathrm{T}
J^{em}_{\mu}(x) J^{em}_{\nu}(0) | \, p \, \rangle.
\end{equation}
Here $J^{em}_{\mu}(x)$ is an electromagnetic current and $| \, p
\, \rangle$ means a proton state. In its turn, an expansion of a
chronological product of two electromagnetic currents near the
light-cone in terms of composite operators $O^{a,\,m}$ looks like:
\begin{eqnarray}\label{004}
&& \mbox{T} J^{em}_{\mu}(x) J^{em}_{\nu}(0) = - g_{\mu\nu}
\sum\limits_{a} \sum\limits_{\mathrm{even} \, m}
C^{(1)}_{a,\,m}(x^2, \mu^2) \, \frac{i^m}{m!} \, x^{\mu_1} \ldots
x^{\mu_m} \, O^{a,\,m}_{\mu_1 \ldots \mu_m}(0;\mu^2)
\nonumber \\
&& + \sum\limits_{a} \sum\limits_{\mathrm{even} \, m}
C^{(2)}_{a,\,m}(x^2, \mu^2) \, \frac{i^m}{m!} \, x^{\mu_1} \ldots
x^{\mu_m} \, O^{a,\,m}_{\mu \nu \mu_1 \ldots \mu_m}(0;\mu^2) +
\ldots \ ,
\end{eqnarray}
where $a = $ {\small NS, S}, $g$. The dots denote other Lorentz
structure, constant term as well as gradient terms. The latter
give no contribution to DIS structure functions. The index $a$
runs all allowed types of the composite operators. The expansion
\eqref{004} is a generalization of the operator product expansion
(OPE) at small distances~\cite{Wilson:69}. It is known as
light-cone OPE~\cite{Frishman:70} (see also
\cite{Buras:82,Yndurain}). The quantity $C^{(i)}_{a,\,m}$ is
called an OPE coefficient function (CF). The composite operators
in Eq.~\eqref{004} need a regularization. After a renormalization
of the operators, there arises a dependence of CF's on a
renormalization scale $\mu$.

On the other hand, the DIS structure functions can be represented
in a factorizable form~\cite{Factorization}. For instance, we have
\begin{equation}\label{006}
\frac{1}{x} \, F_2(x,Q^2) = \sum_a \int\limits_x^1 \frac{dz}{z} \,
C_a \! \left( \! z, \frac{Q^2}{M^2} \right) \, f_a \! \left(
\frac{x}{z}, M^2 \right).
\end{equation}
Here $f_a(x, M^2)$ is a distribution of a parton of type $a$
inside the nucleon, while quantities $C_a (x, Q^2/M^2)$ are known
as DIS coefficient functions. In Eq.~\eqref{006}, $M$ is a
factorization scale. It is usually identified with the
renormalization scale $\mu$. The $m$-th moment of the DIS
coefficient function, $C_a (m, Q^2/\mu^2)$, is identified with the
Fourier transform of the corresponding OPE CF's in \eqref{004}.
The parton distributions $f_a(x, \mu^2)$ are related with matrix
elements of the composite operators between one-nucleon states
(for details, see \cite{Buras:82}).

The coefficient functions $C_a (x, Q^2/M^2)$ were calculated in
perturbative QCD. For example, contributions of the first and
second orders in $\alpha_s$ can be found in
Refs.~\cite{CF:light_quarks} (where light quarks were considered),
and in Refs.~\cite{CF:heavy_quarks} (where both light and heavy
quarks were accounted for). Note that a very definition of DIS
coefficient functions applies to the diagram (perturbative)
technique, not to the operator formalism. Moreover, since nucleon
wave function is yet unknown, one has to deal with diagrams which
describe a lepton scattering off (nonphysical) quark or gluon
off-shell state.

The light-cone OPE for the scalar theory was  studied in
Refs.~\cite{Zavialov} in which rules for calculating CF's were
presented. Note that the T-product of two scalar currents near the
light-cone was defined in term of so-called bi-local light-ray
composite fields \cite{Zavialov}. The local light-cone expansion
can be obtained by performing a Taylor expansion of the non-local
one~\cite{Bordag:80}. The non-local expansion is more general, but
we restrict ourselves to considering local OPE. In
Refs.~\cite{Chetyrkin:82} a problem of finding $n$-loop
contributions to the OPE CF's were reduced to evaluating of
propagator type $(n+1)$-loop Feynman diagrams.

The goal of the present paper is to present a derivation of a
closed representation for the OPE CF's in term of composite
operator Green functions which does not lean on the perturbation
theory. This will be done in the next Section. In
Section~\ref{sec:CF_scalar} we check a validity of our results in
a free scalar field theory. In Section~\ref{sec:CF_QCD} we
calculate nonsinglet CF's in perturbative QCD in order to
demonstrate that our main formulae not only reproduces well-known
expressions for the quark CF's, but enables us to obtain CF's of
the gradient operators in the OPE. The finite renormalization of
CF's is considered in Section~\ref{sec:renorm}. In Appendix~A a
number of useful mathematical formulae is collected. In Appendix~B
we show that our scheme does result in a set of homogeneous
renormalization group equations for the OPE CF's.


\section{OPE coefficient functions and matrix \\
elements of composite operators}
\label{sec:OPE}

Let us define a quark electromagnetic current
\begin{equation}\label{02}
J^{em}_{\mu}(x) =  \overline \Psi (x) \gamma_{\mu} Q \Psi (x),
\end{equation}
where $ \Psi (x)$ is a quark field. The electric charge operator
in \eqref{02},
\begin{equation}\label{04}
Q = \frac{1}{2}(\lambda^3 + \frac{1}{\sqrt{3}} \lambda^8),
\end{equation}
obeys the equations:
\begin{equation}\label{06}
Q^2 =  \frac{1}{6} \left( \lambda^3 + \frac{1}{\sqrt{3}} \lambda^8
+ \frac{4}{3} \lambda^0 \right),
\end{equation}
\begin{equation}\label{08}
\mbox{Sp}(Q^2 \lambda^a) = \frac{1}{3} \left( \delta_{a3} +
\frac{1}{\sqrt{3}} \delta_{a8} + 2 \, \delta_{a0}\right).
\end{equation}
Here $\lambda^a$ ($a = 1,2, \ldots 8$) are the Gell-Mann matrices,
$\mbox{Sp}(\lambda^a \lambda^b) = 2 \delta_{ab}$, and $ \lambda^0$
is the identity matrix.

The operator product expansion (OPE) for the $\mathrm{T}$-product
of two electromagnetic currents looks like (see, for instance,
\cite{Yndurain})
\begin{eqnarray}\label{10}
\mbox{T} J^{em}_{\mu}(x) J^{em}_{\nu}(0) &=& - \frac{1}{6} \,
g_{\mu\nu} \sum\limits_{m=0}^{\infty} \sum\limits_{l=1}^m
C_{1,\,NS}^{m,l}(x^2) \, \frac{i^m}{m!} \, x^{\mu_1} \ldots
x^{\mu_m}
\nonumber \\
&\times& \left[ O^{3,\,m,l}_{NS,\,\mu_1 \ldots \mu_m}(0) +
\frac{1}{\sqrt{3}} O^{8,\,m,l}_{NS,\,\mu_1 \ldots \mu_m}(0)
\right] + \ldots \ ,
\end{eqnarray}
where the dots denote contributions from other Lorentz structures
and singlet quark and gluon operators. It is clear from \eqref{08}
that this expansion should contain nonsinglet (triplet and octet)
and singlet composite operators.

Near the light-cone, a leading contribution comes from twist-2
operators. For instance, quark twist-2 (traceless) operator is of
the form (operator $\mathbf{S}$ means a complete symmetrization in
Lorentz indices):
\begin{eqnarray}\label{12}
O^{a,\,m,l}_{NS,\,\mu_1 \ldots \mu_m}(x) &=& i^{m-1} \mathbf{S} \,
\partial_{\mu_{l+1}} \ldots
\partial_{\mu_m} \overline \Psi (x) \gamma_{\mu_1} D_{\mu_2}
\ldots D_{\mu_l} \lambda^a \Psi (x)
\nonumber \\
&+& (\mbox{terms proportional to } g_{\mu_i\mu_j}).
\end{eqnarray}
Here
\begin{equation}\label{14}
D_{\mu} = \partial_{\mu} + ig \, t_a A^a_{\mu} \qquad (t^a =
\lambda^a/2)
\end{equation}
is a covariant derivative, and  $A^a_{\mu}(x)$ is a gluon field.

If the OPE \eqref{10} is applied to deep inelastic scattering,
only operators of the type
\begin{eqnarray}\label{16}
O^{a,\,m}_{NS,\,\mu_1 \ldots \mu_m}(x) &=& i^{m-1} \mathbf{S} \,
\overline \Psi (x) \gamma_{\mu_1} D_{\mu_2} \ldots D_{\mu_m}
\lambda^a \Psi (x)
\nonumber \\
&+& (\mbox{terms proportional to } g_{\mu_i\mu_j})
\end{eqnarray}
are important, since forward matrix elements of the operators with
$1 \leqslant l \leqslant m-1$ are zero.%
\footnote{It is clear from the relation $\langle \, p \,|\,
\partial_{\mu} \hat{O}\,|\,p' \, \rangle \sim (p - p')_{\mu}$.}
For non-forward Compton scattering, all operators contribute
proportionally to $(p - p')_{\mu_{l+1}} \ldots (p - p')_{\mu_m}$.
The invariant structure which survives at $p' \rightarrow p$ can
be related to the ``skew'' parton distributions. In our notation,
$O^{a,\,m}_{NS,\,\mu_1 \ldots \mu_m} = O^{a,\,m,m}_{NS,\,\mu_1
\ldots \mu_m}$.

In general case, all operators with  $1 \leqslant l \leqslant m$
should be take into account, since they are mixed under the
renormalization, and a renormalized operator
$O^{m,l}_{\mathrm{R}}$ is defined via unrenormalized operators
$O^{m,l}_{\mathrm{U}}$ by the relation (we have dropped
non-relevant indices):
\begin{equation}\label{18}
O^{m,l}_{\mathrm{U}} = \sum_{l'=1}^{l} Z^l_{l'} \,
O^{m,l'}_{\mathrm{R}},
\end{equation}
where $\mathbf{Z}$ is a triangle matrix. In particular, we find
that the composite operator $O^{m,1}$ is multiplicatively
renormalized, while the composite operator $O^{m,m}$, which is
relevant to DIS, is not. In perturbative QCD in the first order of
$\alpha_s = g^2/(4 \pi)$, elements of the matrix $\mathbf{Z}$ are
given by the following expressions~\cite{Yndurain} (with
non-significant finite terms omitted):
\begin{equation}\label{20}
Z_{l l'} =
\left\{%
\begin{array}{cl}
  \displaystyle
    1 + C_F \, \frac{\alpha_s}{4\pi} \, \left[ \frac{1}{\varepsilon}
    + \ln \left( \frac{\bar{\mu}^2}{\mu^2} \right) \right]  \Big[
    - 4 \sum\limits_{j=2}^l \frac{1}{j} -1 + \frac{2}{l(l+1)} \Big], & l' = l,
  \\
  \displaystyle
    C_F \, \frac{\alpha_s}{2\pi} \, \left[ \frac{1}{\varepsilon}
    + \ln \left( \frac{\bar{\mu}^2}{\mu^2} \right) \right] \Big[
    \frac{1}{l-l'} - \frac{1}{l+1}\Big], & l' < l,
  \\
    0, & l' > l,
  \\
\end{array}%
\right.
\end{equation}
where $C_F = (N_c - 1)/2 N_c$, with $N_c$ being a number of
colors. In deriving \eqref{20}, a dimensional
regularization~\cite{Hooft:72} was used, and $\varepsilon = 2 -
D/2$, where $D$ is the number of dimensions. Here and below
$\bar{\mu}$ denotes the \emph{regularization} mass,%
\footnote{The regularization scale $\bar{\mu}$ arises in
dimensional regularization, when one changes an integration
volume, $d^4k \rightarrow \bar{\mu}^{4-D} d^Dk$.}
while $\mu$ denotes the \emph{renormalization} scale.

In what follows, we will be interested in one of nonsinglet quark
operators, namely, in $O^{3,\,m,l}_{NS,\,\mu_1 \ldots \mu_m}(x)$.%
\footnote{It is obviously that the operator
$O^{8,\,m,l}_{NS,\,\mu_1 \ldots \mu_m}(x)$ can be treated in the
same manner. The singlet case is technically more complicated,
since the singlet quark and gluon operators have to be considered
simultaneously.}
Let us introduce brief notations:
\begin{eqnarray}\label{22}
O^{m,l}_{\mu_1 \ldots \mu_m}(x) &\equiv& O^{3,\,m,l}_{NS,\,\mu_1
\ldots \mu_m}(x),
\\
C_{m,l}(x^2) &\equiv& C_{1,\,NS}^{m,l}(x^2).
\end{eqnarray}
After multiplying both sides of Eq.~\eqref{10} by the operator
$O^{n,k}_{\nu_1 \ldots \nu_n}$ (with arbitrary $n$, and $k=1,2
\ldots n$), we get the following relation for
$\mathrm{T}$-products of the composite operators between the
vacuum states:%
\footnote{We used the relation $\mathrm{T} \big( \mathrm{T}
O_1(x_1) \, O_2(x_2) \big) \, O_3(x_3)$ = $\mathrm{T} O_1(x_1) \,
O_2(x_2) \, O_3(x_3)$.}
\begin{eqnarray}\label{24}
&& \int \! d^4x \,\mbox{e}^{iqx} \int \! d^4z \,\mbox{e}^{ipz}
\langle \mathrm{T} J^{em}_{\mu}(x) J^{em}_{\nu}(0) \,
O^{n,k}_{\nu_1 \ldots \nu_n}(z) \rangle
\nonumber \\
&& = - \frac{1}{6} \, g_{\mu\nu} \sum\limits_{m=0}^{\infty}
\sum\limits_{l=1}^m \frac{i^m}{m!} \, \int \! d^4x
\,\mbox{e}^{iqx} \, C_{m,l}(x^2) \, x_{\mu_1} \ldots x_{\mu_m}
\nonumber \\
&& \times \int \! d^4z \,\mbox{e}^{ipz} \langle \mbox{T}
O^{m,l}_{\mu_1 \ldots \mu_m}(0)\,O^{n,k}_{\nu_1 \ldots \nu_n}(z)
\rangle + \ldots \ .
\end{eqnarray}
Here dots mean other Lorentz structures. It is necessary to stress
that propagator of only singlet quark operator has to appeared in
\eqref{24}, due to relation~\eqref{08}.

As usual, we assume that $C_{m,l}(x^2)$ are tempered generalized
functions (this is explicit in perturbative calculations) so the
symbolic relation
\begin{equation}\label{26}
x^{\mu_1} \ldots x^{\mu_m} = (-2i)^m \, \frac{q^{\mu_1} \ldots
q^{\mu_m}}{(-q^2)^m} \, (-q^2)^m \left( \frac{\partial}{\partial
q^2} \right)^m
\end{equation}
holds in connection with the Fourier transform in \eqref{24}. Then
we obtain that
\begin{eqnarray}\label{28}
&& \int \! d^4x \,\mbox{e}^{iqx} \int \! d^4z \,\mbox{e}^{ipz}
\langle \mbox{T} J^{em}_{\mu}(x) J^{em}_{\nu}(0) \, O^{n,k}_{\nu_1
\ldots \nu_n}(z) \rangle
\nonumber \\
&& = - g_{\mu\nu} \sum\limits_{m=0}^{\infty} \sum\limits_{l=1}^m
2^m \, \frac{q^{\mu_1} \ldots q^{\mu_m}}{(-q^2)^m} \,
\tilde{C}^{m,l}(q^2)
\nonumber \\
&& \times \, \! \int \! d^4z \,\mbox{e}^{ipz} \langle \mbox{T}
O^{m,l}_{\mu_1 \ldots \mu_m}(0) \, O^{n,k}_{\nu_1 \ldots \nu_n}(z)
\rangle + \ldots \ ,
\end{eqnarray}
where $\tilde{C}_{m,l}(q^2)$ is a Fourier transform of
$C_{m,l}(x^2)$:%
\footnote{Some authors include the factor $1/m!$ in a definition
of a composite operator.}
\begin{equation}\label{30}
\tilde{C}_{m,l}(q^2) = \frac{1}{m!} \, (-q^2)^m \left(
\frac{\partial}{\partial q^2} \right)^m \int \! d^4x
\,\mbox{e}^{iqx} C_{m,l}(x^2).
\end{equation}

Let $n_\mu$ to be a light-cone 4-vector which is not orthogonal to
4-momentum $p_{\mu}$:
\begin{equation}\label{32}
n_\mu^2 = 0, \qquad p n \neq 0.
\end{equation}
Throughout the paper, we will work in the limit
\begin{equation}\label{34}
p_{\mu}^2 \rightarrow 0, \qquad p_{\mu}^2 < 0.
\end{equation}

Let us now convolute our matrix elements with the projector
\begin{equation}\label{38}
\frac{n^{\nu_1} \ldots n^{\nu_n}}{(pn)^n} \ .
\end{equation}
In particular, we can define the following invariant structure,
\begin{eqnarray}\label{40}
&& \frac{n^{\nu_1} \ldots n^{\nu_n}}{(pn)^n} \int \! d^4x
\,\mbox{e}^{iqx} \int \! d^4z \,\mbox{e}^{ipz} \langle \mbox{T}
J^{em}_{\mu}(x) J^{em}_{\nu}(0) \, O^{n,k}_{\nu_1 \ldots \nu_n}(z)
\rangle
\nonumber \\
&& = - \frac{1}{3} \, g_{\mu\nu} F^{n,k}(\omega, Q^2, p^2) +
\ldots \ ,
\end{eqnarray}
which depends on variables $p^2$,  $Q^2 = - q^2$, and
\begin{equation}\label{42}
\omega = \frac{1}{x} = \frac{2pq}{Q^2}.
\end{equation}
The propagator of the composite operator has the following Lorentz
structure:
\begin{eqnarray}\label{44}
&& \! \int \! d^4z \,\mbox{e}^{ipz} \langle \mbox{T}
O^{m,l}_{\mu_1 \ldots \mu_m}(0) \, O^{n,k}_{\nu_1 \ldots \nu_n}(z)
\rangle
\nonumber \\
&& = 2 \, p_{\mu_1} \ldots p_{\mu_m} p_{\nu_1} \ldots p_{\nu_n}
\langle O^{m,l} O^{n,k} \rangle (p^2)
\nonumber \\
&&  + \, (\mbox{terms proportional to } g_{\mu_i\mu_j}p^2, \
g_{\nu_i\nu_j}p^2, \ g_{\mu_i\nu_j}p^2).
\end{eqnarray}
Equation \eqref{44} means:
\begin{eqnarray}\label{46}
&& \frac{n^{\nu_1} \ldots n^{\nu_n}}{(pn)^n} \! \int \! d^4z
\,\mbox{e}^{ipz} \langle \mbox{T} \, O^{m,l}_{\mu_1 \ldots
\mu_m}(0) O^{n,k}_{\nu_1 \ldots \nu_n}(z) \rangle
\nonumber \\
&& = 2 \, p_{\mu_1} \ldots p_{\mu_m} \langle  O^{m,l} O^{n,k}
\rangle (p^2).
\end{eqnarray}
Note that $F^{n,k}(\omega, Q^2, p^2)$ and $\langle  O^{m,l}
O^{n,k} \rangle(p^2)$ are dimensionless.

Let us note that at $p^2 \rightarrow 0$ propagators of composite
operators of higher twists are suppressed by powers of $p^2$ with
respect to the propagators of twist-2 operators. Thus, our
approach enables us to isolate a contribution from twist-2
operators.

At fixed $Q^2$ and $p^2$, 3-point Green function $\langle \mbox{T}
J^{em}_{\mu} J^{em}_{\nu} \, O^{n,k} \rangle$ has a discontinuity
in the variable $(q+p)^2$ for $(q+p)^2 \geqslant 0$ (that is, for
$\omega \geqslant 1$). By using the unsubtracted dispersion
relation for $F^{n,k}(\omega, Q^2, p^2)$,
\begin{eqnarray}\label{48}
F^{n,k}(\omega, Q^2, p^2) &=& \frac{1}{\pi} \int\limits_1^{\infty}
\! \frac{d \omega'}{\omega' - \omega} \, \mathrm{\rm Im}
F^{n,k}(\omega', q^2, p^2)
\nonumber \\
&=& \frac{1}{2\pi i}  \sum\limits_{m=0}^{\infty} \omega^m
\int\limits_1^{\infty} \! d \omega' \omega'^{-m-1} \,
\mathrm{disc}_{\omega} F^{n,k}(\omega', q^2, p^2),
\end{eqnarray}
one can derive from Eqs.~\eqref{28} and  \eqref{40}, \eqref{44}:
\begin{eqnarray}\label{50}
&& \sum\limits_{l=1}^m \tilde{C}_{m,l}(Q^2) \, \langle  O^{m,l}
O^{n,k} \rangle (p^2) \Big|_{p^2 \rightarrow 0}
\nonumber \\
&& = \frac{1}{2\pi i} \int\limits_0^{1} \! d x x^{m-1}
\mbox{disc}_{(p+q)^2} F^{n,k}(x, Q^2, p^2) \Big|_{p^2 \rightarrow
0}.
\end{eqnarray}
As we will see in the next Sections, both propagator of the
composite operator and matrix element $F^{n,k}$ need a
renormalization already in zero order in strong coupling
constant~\cite{Collins}. Thus, we have $\langle  O^{m,l} O^{n,k}
\rangle = \langle O^{m,l} O^{n,k} \rangle (p^2/\mu^2)$, $F^{n,k} =
F^{n,k} (x, Q^2/p^2, p^2/\mu^2)$, and, consequently,
$\tilde{C}_{m,l} = \tilde{C}_{m,l}(Q^2/\mu^2)$, where $\mu$ is a
\emph{renormalization} scale. Remember that all these quantities
are dimensionless.

The Eq.~\eqref{50} are valid for \emph{all} integer $k \geqslant
1$ (of course, $k \leqslant n$, but we can choose $O^{n,k}$ with
\emph{any} $n$).  Let us put $n \geqslant m$ and define the
matrixes:
\begin{equation}\label{52}
\langle O O \rangle_{lk} = \langle \mathrm{T} O^{m,l} O^{n,k}
\rangle (p^2/\mu^2) \Big|_{p^2 \rightarrow 0},
\end{equation}
and
\begin{equation}\label{54}
\langle J J O \rangle^{k} = \langle \mbox{T} J^{em} J^{em} \,
O^{n,k} \rangle (x, Q^2/p^2 ,p^2/\mu^2) \Big|_{p^2 \rightarrow 0},
\end{equation}
where Lorentz invariant part of 3-point Green function is implied.%
\footnote{There is no dependence of $\langle \! O O \!
\rangle_{lk}$ on indices $m,n$, and no dependence of $\langle \! J
J O \! \rangle^{k}$ on index $n$ except for a trivial factor
$(-1)^n$.}
Then from \eqref{50} we obtain $m$ equations for CF's (for $k=1,2,
\ldots m$):
\begin{equation}\label{56}
\tilde{C}_{m,l} = \sum_{k=1}^m \frac{1}{2\pi i} \int\limits_0^{1}
\! d x x^{m-1} \, \mathrm{disc}_{(p+q)^2}  \langle J J O
\rangle^{k} \, \langle O O \rangle^{-1}_{kl},
\end{equation}
where $\langle O O \rangle^{-1}_{kl}$ is the inverse of the matrix
$\langle O O \rangle_{lk}$.

The formulae~\eqref{50}, \eqref{56} is our main theoretical
result. The equation~\eqref{56} gives an operator definition of
the OPE CF's in term of vacuum matrix elements of the composite
operators.%
\footnote{The electromagnetic current \eqref{02} is a particular
case of a quark composite operator with zero anomalous dimension.}
It is important to stress that our definition does not lean on a
notion of quark distributions.

\section{Coefficient functions in free scalar field \\
theory}
\label{sec:CF_scalar}

As a simple example, let us consider an expansion of T-product of
two composite operators $\phi^2(z) \equiv \,:\!\!\phi(z)
\phi(z)\!\!:$ in free scalar field theory. The scalar fields $\phi
(z)$ are assumed to be real and massless. Using the Wick theorem,
one can find
\begin{equation}\label{60}
\mathrm{T} \phi^2(x) \, \phi^2(y) = [D_c(x-y)]^2  + 2 D_c(x-y)
:\!\phi (y)\phi (x)\!: \, ,
\end{equation}
where
\begin{equation}\label{62}
D_c(z) = \frac{1}{4\pi^2 i}\,\frac{1}{z^2 + i0}
\end{equation}
is a causal propagator of the field $\phi (z)$. By expanding
bilocal operator in powers of variable $(x-y)$ and putting then
$y=0$, we get (here and below a constant term is omitted):
\begin{eqnarray}\label{64}
\mathrm{T} \phi^2(x) \, \phi^2(0) &=&  2 D_c(x)  \sum_{m=0}
\frac{1}{m!} \, x^{\mu_1} x^{\mu_2} \ldots x^{\!\mu_m} \Big[
(-1)^m \, \overleftarrow{\partial}_{\!\mu_1}
\overleftarrow{\partial}_{\!\mu_2} \ldots
\overleftarrow{\partial}_{\!\mu_m}
\nonumber \\
&+& \overrightarrow{\partial}_{\!\mu_1}
\overrightarrow{\partial}_{\!\mu_2} \ldots
\overrightarrow{\partial}_{\!\mu_m} \Big]  \phi^2 (0).
\end{eqnarray}
As usual, the symbol
$\overleftarrow{\partial}_{\!\!\mu}(\overrightarrow{\partial}_{\!\!\mu})$
in \eqref{64} denotes a derivative which acts on the left (right)
standing field $\phi$ in the composite operator $\phi^2 (0)$.

Using an explicit symmetry of  $x^{\mu_1} x^{\mu_2} \ldots
x^{\!\mu_m}$ in indices, one can write
\begin{eqnarray}\label{65}
&& (-1)^m \, x^{\mu_1} \ldots x^{\!\mu_m} \,
\overleftarrow{\partial}_{\!\mu_1}
\overleftarrow{\partial}_{\!\mu_2} \ldots
\overleftarrow{\partial}_{\!\mu_m} = x^{\mu_1} \ldots x^{\!\mu_m}
\nonumber \\
&& \times \sum_{l=0}^m (-1)^l \, {m\choose{l}}
\stackrel{\leftrightarrow}{\partial}_{\mu_{l+1}}
\stackrel{\leftrightarrow}{\partial}_{\mu_{l+2}} \ldots
\stackrel{\leftrightarrow}{\partial}_{\mu_m}
\overrightarrow{\partial}_{\!\mu_1}
\overrightarrow{\partial}_{\!\mu_2} \ldots
\overrightarrow{\partial}_{\!\mu_l},
\end{eqnarray}
where $\stackrel{\leftrightarrow}{\partial}_{\!\mu} =
\overleftarrow{\partial}_{\!\!\mu} +
\overrightarrow{\partial}_{\!\!\mu}$ is a total derivative of
$\phi^2(z)$. Then we obtain from Eqs.~\eqref{64}, \eqref{65}:
\begin{equation}\label{66}
\int \! d^4x \,\mbox{e}^{iqx} \, \mathrm{T} \phi^2(x) \, \phi^2(0)
\, =  \frac{4}{-q^2 + i0} \, \sum_{m=0} \omega^m \sum_{l=0}^m
C_{m,l} \, \phi^2_{m,l}.
\end{equation}
Here $(-q^2 + i0)^{-1}$ is a Fourier transform of $D_c(x)$, and a
brief notation,
\begin{equation}\label{68}
\phi^2_{m,l} \equiv \partial_{\mu_{l+1}} \partial_{\mu_{l+2}}
\ldots \partial_{\mu_m} \phi (0)\, \partial_{\mu_1}
\partial_{\mu_2} \ldots \partial_{\mu_l} \phi (0),
\end{equation}
is introduced. The OPE CF's in Eq.~\eqref{66} are the following
dimensionless quantities:
\begin{eqnarray}\label{70}
C_{m,m} &=& \frac{1}{2} \, [1 + (-1)^m],
\nonumber\\
C_{m,l} &=& \frac{1}{2} \, (-1)^l \, {m\choose{l}}, \qquad l=0,1,
\ldots, m-1.
\end{eqnarray}

\begin{figure}[ht]
\resizebox{\textwidth}{!}{\includegraphics{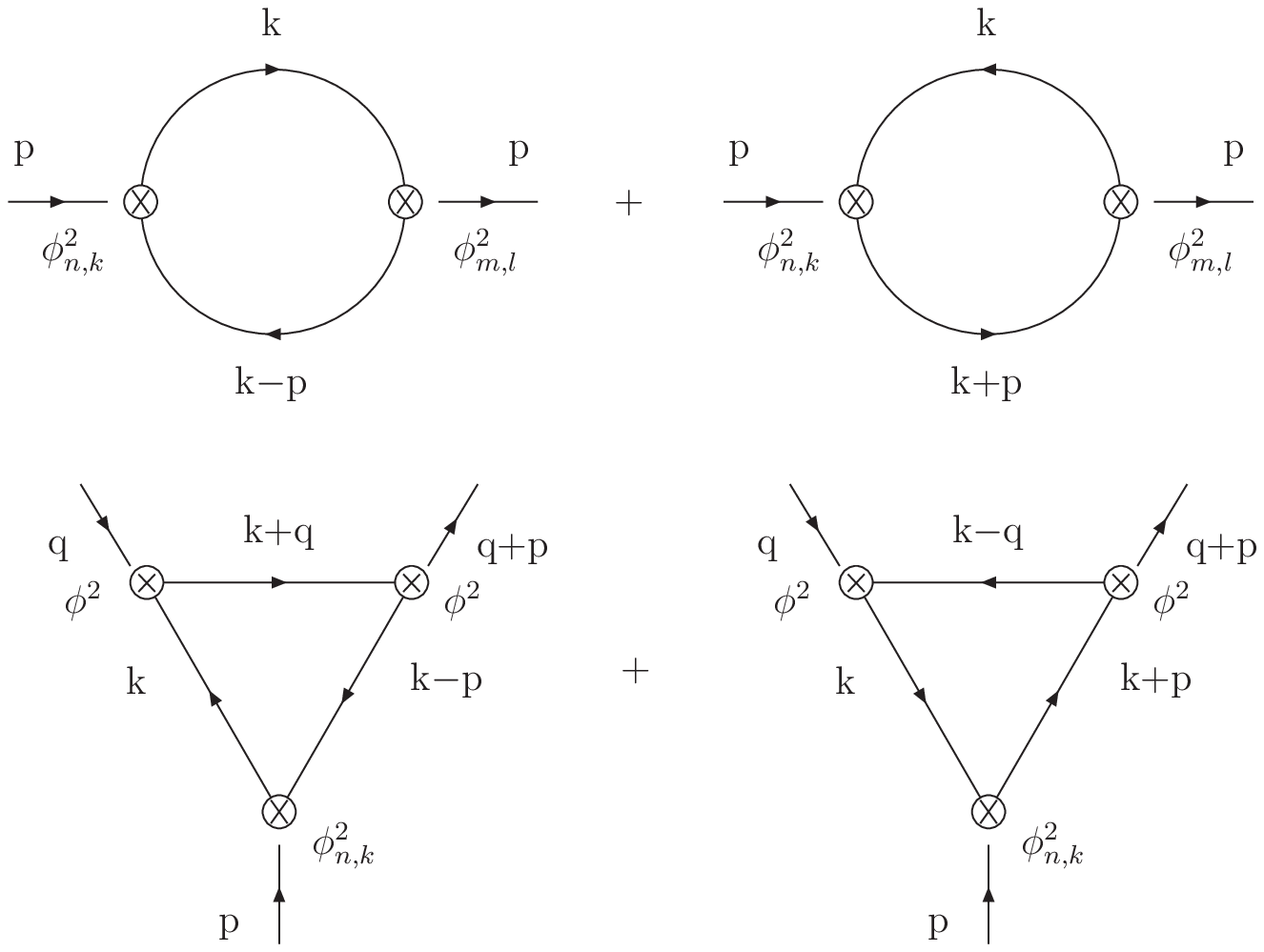}} \caption{The
propagator of the composite operator, $\langle \phi^2_{n,k} \,
\phi^2_{m,l} \rangle$, and the Green function $\langle \phi^2 \,
\phi^2 \, \phi^2_{n,k} \rangle$ in free scalar field theory.}
\label{fig:loop_vertex_0_scalar}
\end{figure}

Let us demonstrate that our main formula~\eqref{50} gives the same
result~\eqref{70}. To do this, one has to calculate both the
propagator of the composite operator, $\langle \phi^2_{m,l} \,
\phi^2_{n,k} \rangle$, and 3-point Green function $\langle \phi^2
\, \phi^2 \, \phi^2_{n,k} \rangle$. The corresponding diagrams are
presented in Fig.~\ref{fig:loop_vertex_0_scalar}. The vertex which
corresponds to the composite operator $\phi^2_{n,k}$ is equal to
$(pn)^{n-k} \, (kn)^k$, with $k$ and $(k-p)$ being ingoing and
outgoing 4-momentum, respectively. The result of our calculations
is the following:
\begin{equation}\label{72}
\langle \phi^2_{m,l} \, \phi^2_{n,k} \rangle (p^2/\bar{\mu}^2)
\Big|_{p^2 \rightarrow 0} = (-1)^{n-1} \frac{1}{16\pi^2} \ln
\left( \frac{\bar{\mu}^2}{-p^2} \right) B(l+1,k+1),
\end{equation}
and
\begin{eqnarray}\label{74}
&& \mathrm{disc}_{(p+q)^2} \, \langle \phi^2 \, \phi^2 \,
\phi^2_{n,k}  \rangle (x, q^2/p^2) \Big|_{p^2 \rightarrow 0}
\nonumber \\
&& = i (-1)^{n-1} \, \frac{1}{4\pi} \, \frac{1}{-q^2 + i0} \, \ln
\left( \frac{q^2}{p^2}\right) \, x \big[ (1 - x)^k + x^k \big].
\end{eqnarray}
The quantity $B(x,y)$ is the beta-function, $B(x,y) = B(y,x)$.

In order to get a finite result for free scalar propagator
\eqref{72}, we had to make a renormalization. As is well known,
any Green function with an insertion of \emph{one} composite
operator is multiplicatively
renormalized~\cite{Zimmermann:73,Itzykson}. The renormalization of
Green functions with the insertion of \emph{two (or more)}
composite operators needs additive counterterms~\cite{Collins}.
The details can be found in Appendix~B.

Form formulae \eqref{72}, \eqref{74} and \eqref{50} (taking into
account the factor in front of the sum in Eq.~\eqref{66}), we
obtain a set of equations for CF's:
\begin{equation}\label{76}
\sum_{l=0}^m C_{m,l} \, B(l+1,k+1) = \frac{1}{2} \, \left[
B(m+1,k+1) + \frac{1}{m+k+1} \right].
\end{equation}
It is easy to check that C-numbers $ C_{m,l}$~\eqref{70} do obey
equations~\eqref{76} for all $m, \, k \geqslant 0$ (for this
purpose, formula~\eqref{A01} from Appendix~A should be used).

\section{Calculations of coefficient functions in \\
perturbative QCD}
\label{sec:CF_QCD}

In this section we will use our formula \eqref{50} for
calculations of the OPE CF's in perturbative QCD. For the
composite operator, we will often use a brief notation $O^{m,l}$
instead of $O^{m,l}_{\mu_1 \ldots \mu_m}(0)$. Contrary to the
matrix element of $O^{m,l}$ between one-particle (quark) states,
$\langle p\,|\,O^{m,l}\,|\,p\rangle$, which defines nonsinglet
quark distributions in DIS, a propagator $\langle O^{m,l} O^{n,k}
\rangle$ is divergent already in zero order in the strong coupling
$\alpha_s$. The matrix element $\langle J\,J \, O^{n,k}
\rangle(Q^2,x,p^2)$ is also divergent, whereas its discontinuity
in variable $(p+q)^2$ is not. The 2-point Green function $\langle
O^{m,l}(x) O^{n,k} (y) \rangle$, is renormalized by a subtraction
of a contact term of the form $(Z^{-1})^l_{l'} \, (Z^{-1})^k_{k'}
\, f_B^{l'k'} \delta (x-y)$, where $Z$ is the renormalization
matrix of the composite operators. The details can be found in
Appendix~B.

We work in the dimensional regularization and use the
$\overline{\mathrm{MS}}$-scheme to renormalize ultra-violet
divergences. Although all calculations will be done in the Feynman
gauge, our results are gauge invariant since we sum all diagrams
in each order of perturbation theory. Remember that in order to
find the OPE CF's, we have to retain only leading terms in the
limit $p^2 \rightarrow 0$. This simplifies our calculations
significantly. We will restrict ourselves by considering leading
terms in $\ln (Q^2/\mu^2)$, although our formula~\eqref{50}
enables one to calculate sub-leading terms as well. In other
words, along with the limit $p^2 \rightarrow 0$, we are interested
in large values of $Q^2$.

\begin{figure}[ht]
\resizebox{\textwidth}{!}{\includegraphics{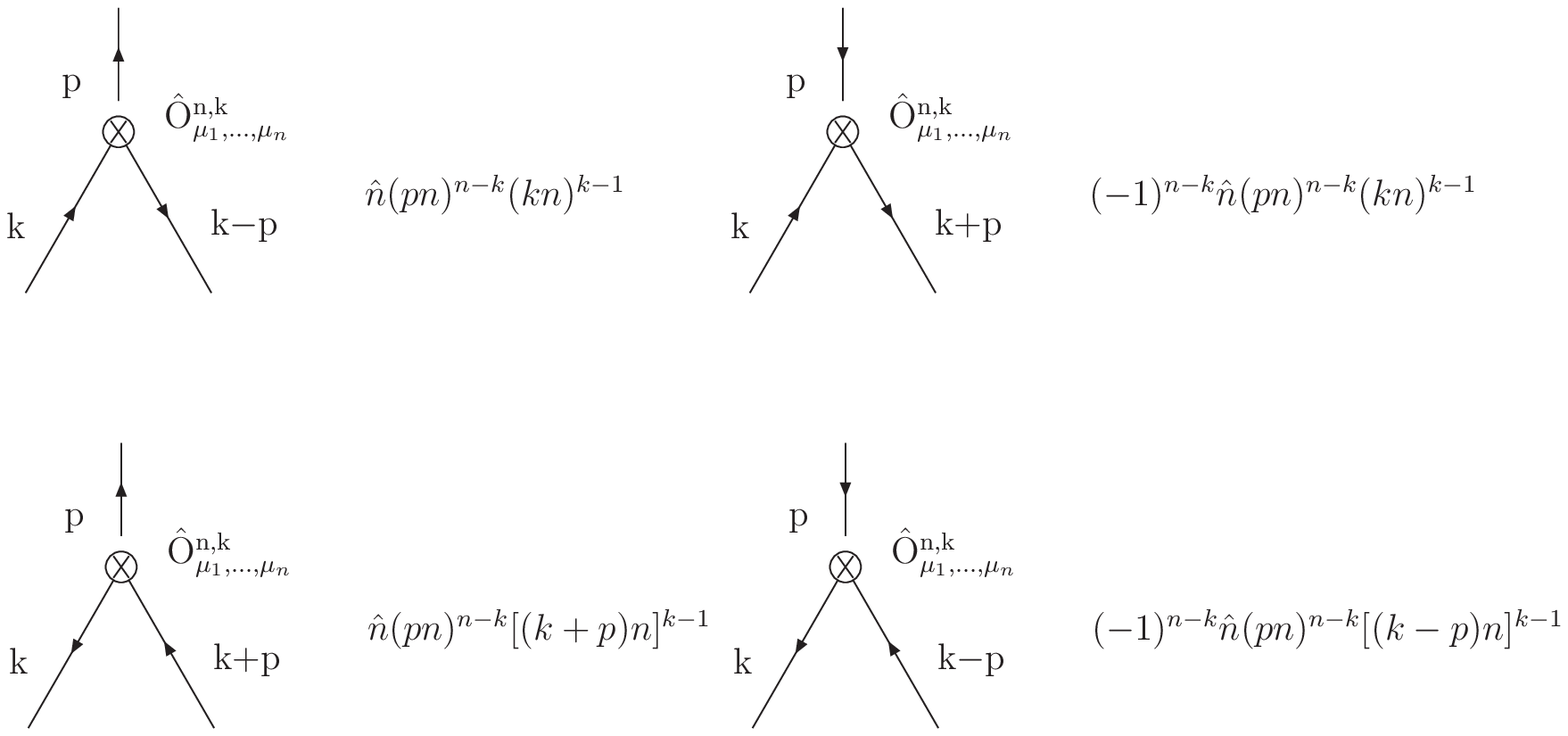}}
\caption{Feynman rules for the composite operators $O^{n,k}_{\mu_1
\ldots \mu_n}$ in the leading (zero) order in strong coupling
$\alpha_s$.} \label{fig:operator_0}
\end{figure}

\begin{figure}[ht]
\resizebox{\textwidth}{!}{\includegraphics{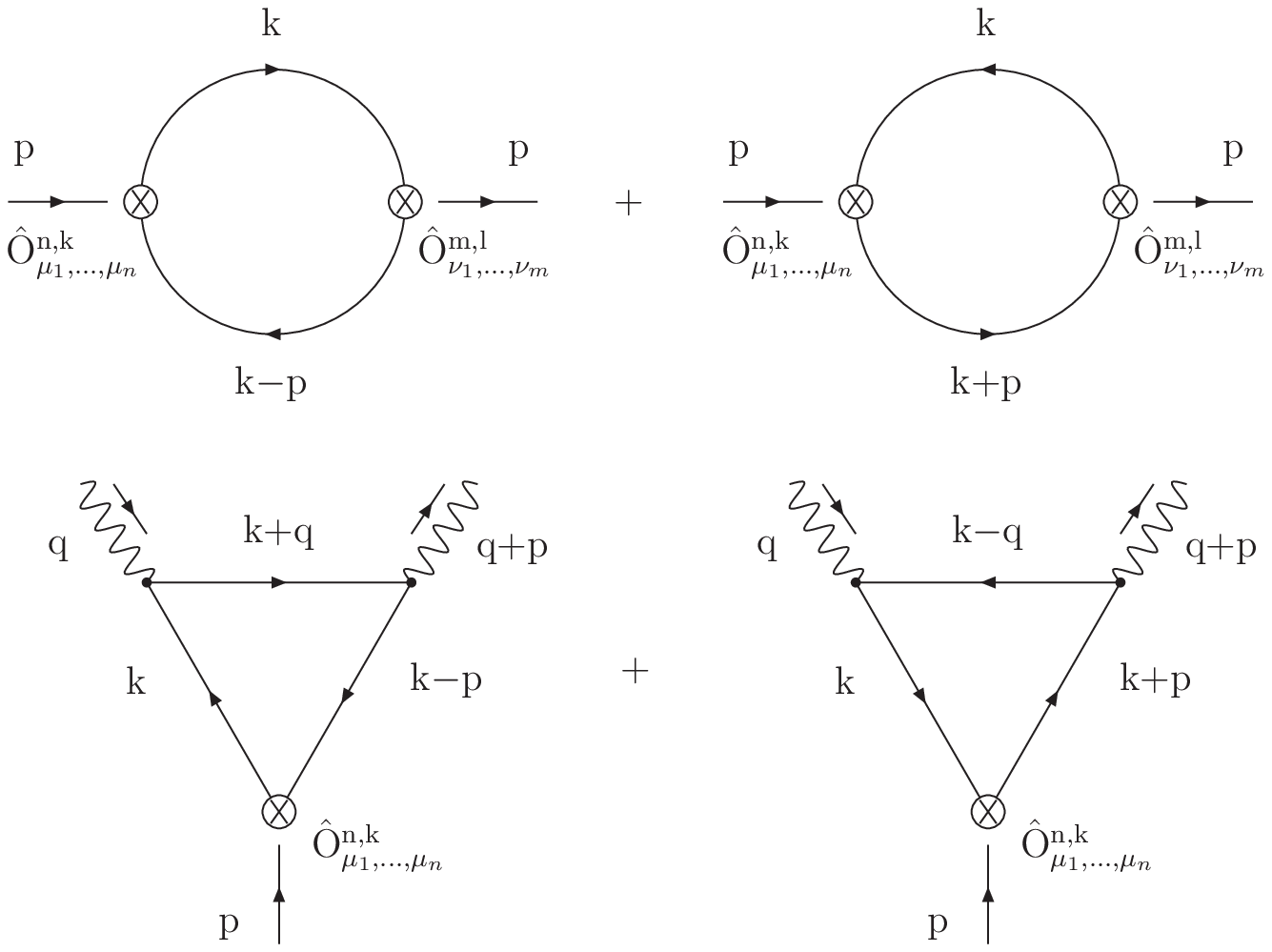}} \caption{The
diagrams for the propagator of the composite operator $\langle
O^{n,k} O^{m,l} \rangle^{(0)}$ and for the matrix element $\langle
J J \, O^{n,k} \rangle^{(0)}$.} \label{fig:loop_vertex_0}
\end{figure}

Let us start from the leading (zero) order in $\alpha_s$.%
\footnote{It corresponds to a theory of free quarks (which
interact only via electromagnetic forces).}
The corresponding Feynman diagrams are shown in
Fig.~\ref{fig:loop_vertex_0}. All diagrams have logarithmic
singularity at $p^2 \rightarrow 0$. They can be easily calculated
with the use of Feynman rules for the composite operators
presented in Fig.~\ref{fig:operator_0}. Note that a vertex which
corresponds to the composite operator is not symmetric with
respect to ingoing and outgoing momenta. For a particular case
with $k=n$, $p=0$, we reproduce the well-known expressions (see
the first paper in Refs.~\cite{Gross:73}).

The result of our calculations in the leading order in the
coupling constant is the following:%
\footnote{Here and in what follows we omit terms which are
sub-leading in the limit $p^2 \rightarrow 0$. The superscript
$^{(n)}$ means that a corresponding quantity is calculated in
$n$-th order in strong coupling constant $\alpha_s$.}
\begin{equation}\label{102}
\langle O^{m,l} O^{n,k} \rangle^{(0)} (p^2/\bar{\mu}^2) =
(-1)^{n-1} \frac{1}{2\pi^2} \ln \left( \frac{\bar{\mu}^2}{-p^2}
\right) B(l+1,k+1),
\end{equation}
and
\begin{eqnarray}\label{104}
&& \mathrm{disc}_{(p+q)^2} \, \langle J\,J \, O^{n,k}
\rangle^{(0)} (x, Q^2/p^2)
\nonumber \\
&& = i (-1)^{n-1} \, \frac{e^2_q}{2\pi} \, \ln \left(
\frac{Q^2}{-p^2}\right) \, \big[ x(1 - x)^{k} - (1 - x)x^{k}
\big],
\end{eqnarray}
where $e^2_q$ is an electric charge of a quark inside a loop. The
terms in square brackets in Eq.~\eqref{104} correspond to two
bottom diagrams in Fig.~\ref{fig:loop_vertex_0} with opposite
directions of the quark momentum inside the loop. As for two top
diagrams in Fig.~\ref{fig:loop_vertex_0}, they are identical. So,
only one of these diagrams was taken into account in
Eq.~\eqref{102}. The formula~\eqref{102} gives a finite (singular
at $p^2 \rightarrow 0$) part of the composite operator propagator
(see our comments after Eq.~\eqref{74}).

Equating factors in front of  $\ln(-p^2)$ in both sides of
Eq.~\eqref{102}, \eqref{104}, and putting $k=1,2, \ldots m$, we
obtain the following set of $m$ equations:
\begin{eqnarray}\label{106}
&& \sum\limits_{l=1}^m \tilde{C}_{m,l}^{(0)} \, B(l+1,k+1) =
\frac{1}{2\pi} \int\limits_0^{1} \! d x \, x^{m} \, (1 - x) [(1 -
x)^{k-1} - x^{k-1}]
\nonumber \\
&& = \frac{1}{2} \left[ B(m+1,k+1) - \frac{1}{(m+k)(m+k+1)}
\right] \, , \quad k=1,2, \ldots m.
\end{eqnarray}
Using formulae \eqref{A02}-\eqref{A08} from Appendix~A, we find
the solution of these equations in the form%
\footnote{Let us stress, we didn't demand  from the very beginning
that $\tilde{C}_{m,m}$ should be equal to zero for odd $m$. It is
a consequence of the fact that electromagnetic interactions
conserve P-parity. Remember that DIS structure function $F_2(x,
Q^2)$ is an even function of variable $x$, and its non-zero
moments are $F_2(n, Q^2) = \tilde{C}_{n,n} \, (Q^2/\mu^2) \langle
\,p\,|\, \hat{O^{n,n}}\,|\,p \,\rangle (\mu^2)$, with $n=2k$.}
\begin{eqnarray}
\tilde{C}_{m,m}^{(0)} &=& \frac{e_q^2}{2} \, [1 + (-1)^m],
\label{107} \\
\tilde{C}_{m,l}^{(0)} &=&  \frac{e_q^2}{2} \, (-1)^l
{m-1\choose{l-1}}, \qquad l=1,2, \ldots, m-1, \label{108}
\end{eqnarray}
where ${n\choose{k}}$ denotes a binomial coefficient. Let us
stress, the quantities $\tilde{C}_{m,l}$ \eqref{108} satisfy
equation \eqref{106} for all integer $k \geqslant 1$, although for
our purpose it was enough to take $m$ values of $k$.

One can check that
\begin{equation}\label{109}
\sum_{l=1}^{m} \, \tilde{C}_{m,l}^{(0)}  \, \frac{1}{(l+1)(l+2)} =
0
\end{equation}
(see Eq.~\eqref{A08}). Note that the RHS of Eq.~\eqref{50} is
identically equal to zero for $k=1$. Indeed, $ \langle J\,J \,
O^{1,1} \rangle (x, Q^2/p^2) = 0$ due to the Furry theorem. On the
other hand, Eq.~\eqref{102} means that $\langle O^{1,1} O^{m,l}
\rangle^{(0)} \sim [(l+1)(l+2)]^{-1}$. We conclude, it is the
relation \eqref{109} due to which Eq.~\eqref{50} is satisfied in
the leading order in $\alpha_s$. As we will see below, a similar
relation is valid for $\tilde{C}_{m,l}^{(1)}$ as well.

\begin{figure}[ht]
\resizebox{\textwidth}{!}{\includegraphics{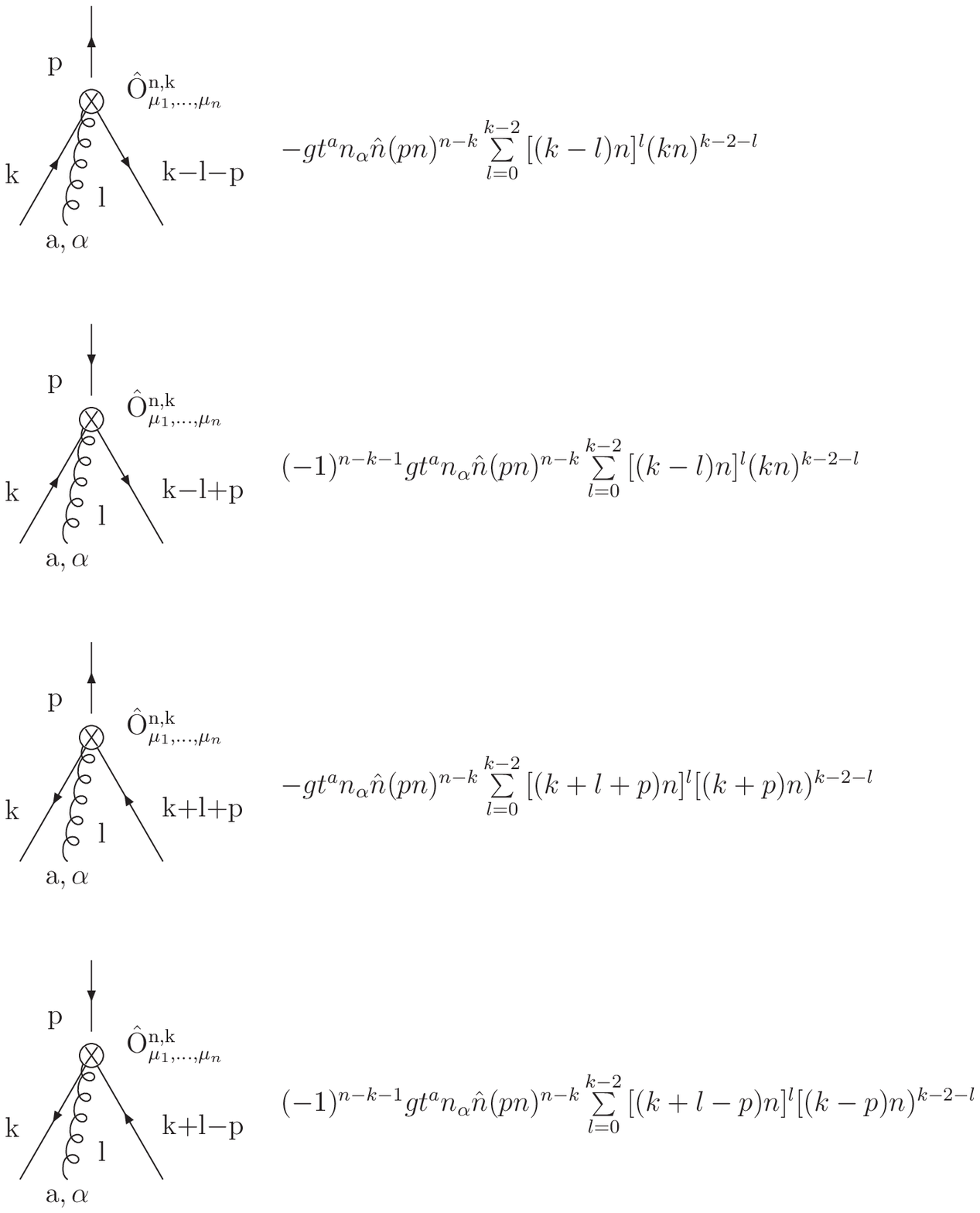}}
\caption{Feynman rules for the composite operators $O^{n,k}_{\mu_1
\ldots \mu_n}$ in the first order in strong coupling $\alpha_s$.}
\label{fig:operator_1}
\end{figure}

Now let us consider the next order in $\alpha_s$. The diagrams
describing propagator of the composite operator in this order are
presented in Figs.~\ref{fig:loop_1_a}, \ref{fig:loop_1_b} and
\ref{fig:loop_1_c}. Using Feynman rules for the composite
operators shown in Fig.~\ref{fig:operator_1}, we obtain the
following contribution of the diagrams in Fig.~\ref{fig:loop_1_a}:
\begin{eqnarray}\label{110}
&& \langle O^{n,k} O^{m,l} \rangle^{(1)}_a (p^2/\mu^2) =
(-1)^{n-1} \frac{\alpha_s}{8 \pi^3} \, C_F \ln \left(
\frac{\bar{\mu}^2}{-p^2} \right) \, \ln \left( \frac{\mu^2}{-p^2}
\right)
\nonumber \\
&& \times \Bigg\{ \, \frac{2}{k(k + 1)} \, \Bigg[ B(l+1,k+1) +
\sum_{j=1}^{k-1} B(l+1,j+1) \Bigg]
\nonumber \\
&& - B(l+1,k+1) + (k \rightleftarrows l) \Bigg\},
\end{eqnarray}
where $C_F = (N_c^2 - 1)/2N_c$, with $N_c$ being a number of
colors. For the diagrams in  Fig.~\ref{fig:loop_1_b}, we get the
expression:
\begin{eqnarray}\label{112}
&& \langle O^{n,k} O^{m,l} \rangle^{(1)}_b (p^2/\mu^2) =
(-1)^{n-1} \frac{\alpha_s}{4 \pi^3} \, C_F \ln \left(
\frac{\bar{\mu}^2}{-p^2} \right) \, \ln \left( \frac{\mu^2}{-p^2}
\right)
\nonumber \\
&& \times \left[ \left(-2 \sum_{j=2}^k \frac{1}{j} \right)
B(l+1,k+1) \right.
\nonumber \\
&& +  \left. \sum_{j=1}^{k-1} \left( \frac{1}{k-j} - \frac{1}{k}
\right) B(l+1,j+1) + (k \rightleftarrows l) \right]
\end{eqnarray}
Both diagrams in Fig.~\ref{fig:loop_1_c} are equal to zero since
they are proportional to $n_{\mu}^2$. Thus, the sum of diagrams
which make a contribution to the propagator of the composite
operator is given by
\begin{eqnarray}\label{114}
&& \langle O^{n,k} O^{m,l} \rangle^{(1)} (p^2/\mu^2) = (-1)^{n-1}
\frac{\alpha_s}{8 \pi^3} \, C_F \ln \left(
\frac{\bar{\mu}^2}{-p^2} \right) \, \ln \left( \frac{\mu^2}{-p^2}
\right)
\nonumber \\
&& \times \left[ \left(-4 \sum_{j=2}^k \frac{1}{j} -1 +
\frac{2}{k(k + 1)} \right) \right. B(l+1,k+1)
\nonumber \\
&& + \, 2 \, \left. \sum_{j=1}^{k-1} \left( \frac{1}{k-j} -
\frac{1}{k+1} \right) B(l+1,j+1)  + (k \rightleftarrows l)
\right].
\end{eqnarray}
Note, that $\langle O^{n,k} O^{m,l} \rangle^{(1)} = 0$ for $k=1$
or $l=1$. Equation~\eqref{114} can be presented in the form:
\begin{eqnarray}\label{115}
&& \langle O^{n,k} O^{m,l} \rangle^{(1)} (p^2/\mu^2) =
\frac{\alpha_s}{4 \pi} \, C_F \ln \left( \frac{\mu^2}{-p^2}
\right)
\nonumber \\
&& \times \left[ \langle O^{n,k} O^{m,l} \rangle^{(0)}
(p^2/\bar{\mu}^2) \left(-4 \sum_{j=2}^k \frac{1}{j} -1 +
\frac{2}{k(k + 1)} \right) \right.
\nonumber \\
&& +  \left. 2  \sum_{j=1}^{k-1} \left( \frac{1}{k-j} -
\frac{1}{k+1} \right) \, \langle O^{n,j} O^{m,l} \rangle^{(0)}
(p^2/\bar{\mu}^2)  + (k \rightleftarrows l) \right].
\end{eqnarray}

For our further purposes, it is useful to present \eqref{114} in a
form which has no explicit symmetry in $l$ and $k$:
\begin{eqnarray}\label{116}
&& \langle O^{n,k} O^{m,l} \rangle^{(1)} (p^2/\mu^2) = (-1)^{n-1}
\frac{\alpha_s}{4 \pi^3} \, C_F \ln \left(
\frac{\bar{\mu}^2}{-p^2} \right) \, \ln \left( \frac{\mu^2}{-p^2}
\right)
\nonumber \\
&& \times \left[ \left(-4 \sum_{j=2}^k \frac{1}{j} -1 +
\frac{2}{k(k + 1)} \right) \right. B(l+1,k+1)
\nonumber \\
&& + \, 2 \left. \sum_{j=1}^{k-1} \left( \frac{1}{k-j} -
\frac{1}{k+1} \right) B(l+1,j+1) \right].
\end{eqnarray}
In deriving Eq.~\eqref{116} from Eq.~\eqref{114}, the formulae
from the Appendix~A were used. Taking into account \eqref{108}, we
find
\begin{eqnarray}\label{118}
&& \sum_{l=1}^{m} \, \tilde{C}_{m,l}^{(0)} \, \langle O^{n,k}
O^{m,l} \rangle^{(1)} (p^2/\mu^2) = (-1)^{n-1} \, e_q^2 \,
\frac{\alpha_s}{8 \pi^3} \, C_F \ln \left(
\frac{\bar{\mu}^2}{-p^2} \right) \, \ln \left( \frac{\mu^2}{-p^2}
\right)
\nonumber \\
&& \times \left\{ [1 + (-1)^m]\left[ \left( -4 \sum_{j=2}^k
\frac{1}{j}  - 1 + \frac{2}{k(k+1)} \right) B(m+1,k+1) \right.
\right.
\nonumber \\
&& + \left. 2 \, \sum_{j=1}^{k-1} \left( \frac{1}{k-j} -
\frac{1}{k+1} \right) B(m+1,j+1) \right]
\nonumber \\
&& + \sum_{l=1}^{m-1} (-1)^l {m-1\choose{l-1}} \left[ \left( -4
\sum_{j=2}^k \frac{1}{j}  - 1 + \frac{2}{k(k+1)} \right)
B(l+1,k+1) \right.
\nonumber \\
&& + \left. \left. 2 \, \sum_{j=1}^{k-1} \left( \frac{1}{k-j} -
\frac{1}{k+1} \right) B(l+1,j+1) \right] \right\}.
\end{eqnarray}
As one can see from \eqref{118}, $\sum_{l=1}^{m} \,
\tilde{C}_{m,l}^{(0)} \, \langle O^{n,k} O^{m,l} \rangle^{(1)} =
0$ for $k=1$ and $m=1$.

The expression~\eqref{118} can be simplified and presented as%
\footnote{We used the relation $B(m+1,2) - [(m+1)(m+2)]^{-1} =
0$.}
\begin{eqnarray}\label{120}
&& \sum_{l=1}^{m} \, \tilde{C}_{m,l}^{(0)} \, \langle O^{n,k}
O^{m,l} \rangle^{(1)} (p^2/\mu^2) = (-1)^{n-1} \, e_q^2 \,
\frac{\alpha_s}{8 \pi^3} \, C_F \ln \left(
\frac{\bar{\mu}^2}{-p^2} \right) \, \ln \left( \frac{\mu^2}{-p^2}
\right)
\nonumber \\
&& \times \left\{ \left( -4 \sum_{j=2}^k \frac{1}{j}  - 1 +
\frac{2}{k} + \frac{2}{m} \right) \right. \left[ B(m+1,k+1) -
\frac{1}{(m+k)(m+k+1)} \right]
\nonumber \\
&& + \, 2 \, \sum_{j=2}^{k-1} \frac{1}{k-j} \left[ B(m+1,j+1) -
\frac{1}{(m+j)(m+j+1)} \right]
\nonumber \\
&& + \left. \frac{2(m-1)(k-1)}{m(m+1)(k+1)(m+k)} \right\}.
\end{eqnarray}
With the use of Eq.~\eqref{A30} from Appendix~A, one can rewrite
\eqref{120} in the form which has an explicit symmetry in $k, m$:
\begin{eqnarray}\label{122}
&& \sum_{l=1}^{m} \, \tilde{C}_{m,l}^{(0)} \, \langle O^{n,k}
O^{m,l} \rangle^{(1)} (p^2/\mu^2) = (-1)^{n-1} \, e_q^2 \,
\frac{\alpha_s}{8 \pi^3} \, C_F \ln \left(
\frac{\bar{\mu}^2}{-p^2} \right) \, \ln \left( \frac{\mu^2}{-p^2}
\right)
\nonumber \\
&& \times \left\{ 2 \left( - \sum_{j=2}^k \frac{1}{j}  -
\sum_{j=2}^m \frac{1}{j}  + 1 + \frac{2}{k} + \frac{2}{m} \right)
\right.
\nonumber \\
&& \times \left[ B(m+1,k+1) - \frac{1}{(m+k)(m+k+1)} \right]
\nonumber \\
&& + \left[ \sum_{j=1}^{k-1} \frac{1}{k-j} \, B(m+1,j+1) +
\sum_{j=1}^{m-1} \frac{1}{m-j} \, B(k+1,j+1) \right]
\nonumber \\
&& - \left. \frac{2}{(m+k)(m+k+1)} \sum_{j=2}^{m+k-1} \frac{1}{j}
\, + \, \frac{2 -(m+k)}{k(k+1)m(m+1)} \right\}.
\end{eqnarray}

Now let us calculate the Green function which contains one
composite operator and two electromagnetic currents. The diagrams
in Fig.~\ref{fig:vertex_1_a} give
\begin{eqnarray}\label{124}
&& \mathrm{disc}_{(p+q)^2} \, \langle J\,J \, O^{n,k}
\rangle^{(1)}_a (x, Q^2/p^2 ,p^2/\mu^2)
\nonumber \\
&& = i(-1)^{n-1} \, e_q^2 \, \frac{\alpha_s}{8 \pi^2} \, C_F \ln
\left( \frac{Q^2}{-p^2} \right) \, \ln \left( \frac{\mu^2}{-p^2}
\right)
\nonumber \\
&& \times \Bigg\{ \Bigg[ \frac{2}{k(k + 1)} - 1 \Bigg] \, \Big[
x(1 - x)^{k} - (1 - x)x^{k} \Big]
\nonumber \\
&& + \frac{2}{k(k + 1)} \, \sum_{j=2}^{k-1} \, \Big[ x(1 - x)^{j}
- (1 - x)x^{j} \Big] \Bigg\}.
\end{eqnarray}
The equation~\eqref{124} can be rewritten as follows:
\begin{eqnarray}\label{125}
&& \mathrm{disc}_{(p+q)^2} \, \langle J\,J \, O^{n,k}
\rangle^{(1)}_a (x, Q^2/p^2 ,p^2/\mu^2) = \frac{\alpha_s}{4 \pi}
\, C_F  \ln \left( \frac{\mu^2}{-p^2} \right)
\nonumber \\
&& \times \Bigg\{ \Bigg[ \frac{2}{k(k + 1)} - 1 \Bigg] \,
\mathrm{disc}_{(p+q)^2} \, \langle J\,J \, O^{n,k} \rangle^{(0)}
(x, Q^2/p^2)
\nonumber \\
&& +  \frac{2}{k(k + 1)} \, \sum_{j=1}^{k-1} \,
\mathrm{disc}_{(p+q)^2} \, \langle J\,J \, O^{n,j} \rangle^{(0)}
(x, Q^2/p^2) \Bigg\},
\end{eqnarray}
with zero order expression given by \eqref{104}.

The contribution from the sum of the diagrams in
Fig.~\ref{fig:vertex_1_b} is equal to
\begin{eqnarray}\label{126}
&& \mathrm{disc}_{(p+q)^2} \, \langle J\,J \, O^{n,k}
\rangle^{(1)}_b (x, Q^2/p^2, p^2/\mu^2)
\nonumber \\
&& = i(-1)^{n-1} \, e_q^2 \, \frac{\alpha_s}{4 \pi^2} \, C_F \ln
\left( \frac{Q^2}{-p^2} \right) \, \ln \left( \frac{\mu^2}{-p^2}
\right)
\nonumber \\
&& \times \left\{ -2 \sum_{j=2}^k \frac{1}{j}  \,\, \Big[ x(1 -
x)^{k} - (1 - x)x^{k} \Big] \right.
\nonumber \\
&& + \left . \, \sum_{j=2}^{k-1} \left( \frac{1}{k-j} -
\frac{1}{k} \right) \, \Big[ x(1 - x)^{j} - (1 - x)x^{j} \Big]
\right\}.
\end{eqnarray}
Note again that Eq.~\eqref{126} can be rewritten in terms of an
absorptive part of the 3-point Green function calculated in the
leading order in $\alpha_s$:
\begin{eqnarray}\label{127}
&& \mathrm{disc}_{(p+q)^2} \, \langle J\,J \, O^{n,k}
\rangle^{(1)}_b (x, Q^2/p^2, p^2/\mu^2) = \frac{\alpha_s}{2 \pi}
\, C_F  \ln \left( \frac{\mu^2}{-p^2} \right)
\nonumber \\
&& \times \left\{ \left( -2 \sum_{j=2}^k \frac{1}{j} \right)
\mathrm{disc}_{(p+q)^2} \, \langle J\,J \, O^{n,k} \rangle^{(0)}
(x, Q^2/p^2) \right.
\nonumber \\
&& + \left . \sum_{j=1}^{k-1} \left( \frac{1}{k-j} - \frac{1}{k}
\right) \mathrm{disc}_{(p+q)^2} \, \langle J\,J \, O^{n,j}
\rangle^{(0)} (x, Q^2/p^2) \right\}.
\end{eqnarray}

The diagram in Fig.~\ref{fig:vertex_1_c} is sub-leading at $p^2
\rightarrow 0$, since it has no singularities at $p_{\mu} = 0$.
The diagrams in Figs.~\ref{fig:current_1_a} and
\ref{fig:current_1_b} describe a renormalization of the
electromagnetic current $J^{em}$ which is conserved and,
therefore, has no anomalous dimensions. As a result, these
diagrams are suppressed by the inverse of $\ln (\mu^2/(-p^2))$
with respect to \eqref{124} and \eqref{126}. Thus, we have to sum
only expressions \eqref{124} and \eqref{126}:
\begin{eqnarray}\label{128}
&& \frac{1}{2\pi} \int\limits_0^{1} \! d x x^{m-1} \,
\mathrm{disc}_{(p+q)^2} \, \langle J\,J \, O^{n,k} \rangle^{(1)}
(x, Q^2/p^2, p^2/\mu^2)
\nonumber \\
&& = i (-1)^{n-1} \, e_q^2 \, \frac{\alpha_s}{8 \pi^3} \, C_F \ln
\left( \frac{Q^2}{-p^2} \right) \, \ln \left( \frac{\mu^2}{-p^2}
\right)
\nonumber \\
&& \times \left\{ \left( -4 \sum_{j=2}^k \frac{1}{j} \, - 1 +
\frac{2}{k} \right) \right. \left[ B(m+1,k+1) -
\frac{1}{(m+k)(m+k+1)} \right]
\nonumber \\
&& + \, 2 \, \sum_{j=2}^{k-1} \frac{1}{k-j} \left[ B(m+1,j+1) -
\frac{1}{(m+j)(m+j+1)} \right]
\nonumber \\
&& - \, \frac{2}{k+1} \left[ B(m+1,k+1) - \frac{1}{(m+k)(m+k+1)}
\right]
\nonumber \\
&& - \left. \frac{2}{k+1} \sum_{j=2}^{k-1} \left[ B(m+1,j+1) -
\frac{1}{(m+j)(m+j+1)} \right] \right\}.
\end{eqnarray}
Making use of equation~\eqref{A12}, one can easily calculate the
sum in the last line of equation \eqref{128} and rewrite it in the
form:
\begin{eqnarray}\label{130}
&& \frac{1}{2\pi} \int\limits_0^{1} \! d x x^{m-1} \,
\mathrm{disc}_{(p+q)^2} \, \langle J\,J \, O^{n,k} \rangle^{(1)}
(x, Q^2/p^2, p^2/\mu^2)
\nonumber \\
&& = i (-1)^{n-1} \, e_q^2 \, \frac{\alpha_s}{8 \pi^3} \, C_F \ln
\left( \frac{Q^2}{-p^2} \right) \, \ln \left( \frac{\mu^2}{-p^2}
\right)
\nonumber \\
&& \times \left\{ \left( -4 \sum_{j=2}^k \frac{1}{j}  - 1 +
\frac{2}{k} + \frac{2}{m} \right) \right. \left[ B(m+1,k+1) -
\frac{1}{(m+k)(m+k+1)} \right]
\nonumber \\
&& + \, 2 \, \sum_{j=2}^{k-1} \frac{1}{k-j} \left[ B(m+1,j+1) -
\frac{1}{(m+j)(m+j+1)} \right]
\nonumber \\
&& + \frac{2(m-1)(k-1)}{m(m+1)(k+1)(m+k)} \Bigg\}.
\end{eqnarray}

The set of equations for the OPE CF's
$\tilde{C}_{m,l}^{(1)}(Q^2/\mu^2)$ ($l = 1,2, \ldots m$) looks
like:
\begin{eqnarray}\label{132}
&& \sum_{l=1}^{m} \, \tilde{C}_{m,l}^{(1)} (Q^2/\mu^2) \, \langle
O^{n,k} O^{m,l} \rangle^{(0)} (p^2) + \sum_{l=1}^{m} \,
\tilde{C}_{m,l}^{(0)} \, \langle O^{n,k} O^{m,l} \rangle^{(1)}
(p^2)
\nonumber \\
&& = \frac{1}{2\pi i} \int\limits_0^{1} \! d x x^{m-1} \,
\mathrm{disc}_{(p+q)^2} \, \langle J\,J \, O^{n,k} \rangle^{(1)}
(x, Q^2/p^2, p^2/\mu^2).
\end{eqnarray}
From \eqref{132} and \eqref{120}, \eqref{130} one, therefore, can
get $m$ equations,
\begin{eqnarray}\label{134}
&& \sum_{l=1}^{m} \, \tilde{C}_{m,l}^{(1)} (Q^2/\mu^2) \,
B(l+1,k+1) = \, e_q^2 \, \frac{\alpha_s}{8 \pi} \, C_F \ln \left(
\frac{Q^2}{\mu^2} \right)
\nonumber \\
&& \times \left\{ \left( -4 \sum_{j=2}^k \frac{1}{j}  - 1 +
\frac{2}{k} + \frac{2}{m} \right) \right. \left[ B(m+1,k+1) -
\frac{1}{(m+k)(m+k+1)} \right]
\nonumber \\
&& + \, 2 \, \sum_{j=2}^{k-1} \frac{1}{k-j} \left[ B(m+1,j+1) -
\frac{1}{(m+j)(m+j+1)} \right]
\nonumber \\
&& +  \frac{2(m-1)(k-1)}{m(m+1)(k+1)(m+k)} \Bigg\},
\end{eqnarray}
for $m$ different values of $k = 1, 2 \ldots m$. In particular, we
obtain for $k=1$:
\begin{equation}\label{135}
\sum_{l=1}^{m} \, \tilde{C}_{m,l}^{(1)} (Q^2/\mu^2) \,
\frac{1}{(l+1)(l+2)} = 0.
\end{equation}
This equation is similar to formula~\eqref{109} which was obtained
in the leading order.

However, in order to find $\tilde{C}_{m,l}^{(1)}$, it is much more
convenient to use equivalent form of the expression in the RHS of
Eq.~\eqref{134} For this purpose, one should compare \eqref{120}
with \eqref{118}, and then exploit the above demonstrated symmetry
in $k$, $l(k, m)$~\eqref{122}:
\begin{eqnarray}\label{136}
&& \sum_{l=1}^{m} \, \tilde{C}_{m,l}^{(1)} (Q^2/\mu^2) \,
B(k+1,l+1) = \, e_q^2 \, \frac{\alpha_s}{8 \pi} \, C_F \ln \left(
\frac{Q^2}{\mu^2} \right)
\nonumber \\
&& \times \left\{ [1 + (-1)^m]\left[ \left( -4 \sum_{j=2}^m
\frac{1}{j}  - 1 + \frac{2}{m(m+1)} \right) B(k+1,m+1) \right.
\right.
\nonumber \\
&& + \left. 2 \, \sum_{j=1}^{m-1} \left( \frac{1}{m-j} -
\frac{1}{m+1} \right) B(k+1,j+1) \right]
\nonumber \\
&& + \sum_{l=1}^{m-1} (-1)^l {m-1\choose{l-1}} \left[ \left( -4
\sum_{j=2}^l \frac{1}{j}  - 1 + \frac{2}{l(l+1)} \right)
B(k+1,l+1) \right.
\nonumber \\
&& + \left. \left. 2 \, \sum_{j=1}^{l-1} \left( \frac{1}{l-j} -
\frac{1}{l+1} \right) B(k+1,j+1) \right] \right\}.
\end{eqnarray}
Thus, we get $m$ equations (corresponding to $k = 1, 2 \ldots m$)
for $m$ coefficient functions. The solution of these
equations~\eqref{136} is rather easy to find, the result is
\begin{eqnarray}\label{138}
&& \tilde{C}_{m,m}^{(1)}(Q^2/\mu^2) = e_q^2 \, \frac{\alpha_s}{8
\pi} \, C_F \ln \left( \frac{Q^2}{\mu^2} \right) \,
\nonumber \\
&& \times [1 + (-1)^m] \left[ -4 \sum_{j=2}^m \frac{1}{j} - 1 +
\frac{2}{m(m+1)} \right],
\end{eqnarray}
and
\begin{eqnarray}\label{140}
&& \tilde{C}_{m,l}^{(1)} (Q^2/\mu^2) = e_q^2 \, \frac{\alpha_s}{4
\pi} \, C_F \ln \left( \frac{Q^2}{\mu^2} \right)
\nonumber \\
&& \times \left\{ \frac{1}{2} (-1)^l {m-1\choose{l-1}} \left[ -4
\sum_{j=2}^l  \frac{1}{j}  - 1 + \frac{2}{l(l+1)} \right] \right.
\nonumber \\
&& + \left( \frac{1}{m-l} - \frac{1}{m+1} \right)
\nonumber \\
&& + \left. \sum_{k=l+1}^m (-1)^k {m-1\choose{k-1}} \left(
\frac{1}{k-l} - \frac{1}{k+1} \right) \right\},
\end{eqnarray}
for $l=1,2, \ldots, m-1$.%
\footnote{It easy to check that our CF's~\eqref{138}, \eqref{140}
satisfy Eqs.~\eqref{136} not only for $k = 1, 2 \ldots m$, but
also for arbitrary integer $k \geqslant 1$.}
We used formulae for summation in $l$ (\eqref{A08}, \eqref{A10})
and $j$ (\eqref{A22}) from Appendix~A. In particular, we get from
\eqref{140} ($m \geqslant 2$):
\begin{equation}\label{141}
\tilde{C}_{m, 1}^{(1)} = e_q^2 \, \frac{\alpha_s}{4 \pi} \, C_F
\ln \left( \frac{Q^2}{\mu^2} \right) \Bigg[ \sum_{j=1}^m
\frac{1}{j}  - \frac{1}{2} + \frac{1}{m-1} - \frac{2}{m+1} \Bigg].
\end{equation}

As one can see from \eqref{138}, ``major'' coefficient function
$\tilde{C}_{m, m}^{(1)}$ is defined by well-known anomalous
dimension of $O^{m, m}$~\cite{Yndurain},
\begin{equation}\label{142}
\gamma^{_{NS}}_m = \frac{\alpha_s}{2 \pi} \, C_F \, \left[ 4
\sum_{j=2}^m \frac{1}{j} + 1 - \frac{2}{m(m+1)} \right],
\end{equation}
in spite of the fact that $O^{m, m}$ mixes with all operators
$O^{m, l}$ ($l = 1, 2. \ldots m$) under the renormalization. We
have reproduced the standard expression for the coefficient
function $\tilde{C}_{m, m}(Q^2/\mu^2)$, and, simultaneously, have
calculated ``gradient'' OPE coefficient functions $\tilde{C}_{m,
l}(Q^2/\mu^2)$ ($l = 1, 2. \ldots m-1$) in zero and first order in
$\alpha_s$ (see Eqs.~\eqref{108} and \eqref{140}).

Let us emphasize that our operator definition of the coefficient
functions $C^{m,l}(Q^2/\mu^2)$~\eqref{56} results in a homogeneous
renormalization group equation for $C^{m,l}(Q^2/\mu^2)$ with
respect to $\mu$, in spite of the fact that composite operator
Green functions $\langle O^{n,k} O^{m,l} \rangle$ need an
additional renormalization. The details are discussed in
Appendix~B.

\section{Finite renormalization of composite \\
operators and rescaling of coefficient \\
functions}
\label{sec:renorm}

With accounting for equations \eqref{107}, \eqref{108} and
\eqref{138}, \eqref{140}, the expressions for the OPE CF's can be
rewritten in the following form:
\begin{eqnarray}\label{202}
\tilde{C}_{m,m}(Q^2/\mu^2) &=& \tilde{C}_{m,m}^{(0)} \, \Bigg\{ 1
+ \frac{\alpha_s}{4 \pi} \, C_F \ln \left( \frac{Q^2}{\mu^2}
\right)
\nonumber \\
&\times&  \left[ -4 \sum_{j=2}^m \frac{1}{j} - 1 +
\frac{2}{m(m+1)} \right] \Bigg\},
\end{eqnarray}
\begin{eqnarray}\label{204}
\tilde{C}_{m,l} (Q^2/\mu^2) &=& \tilde{C}_{m,l}^{(0)} \, \Bigg\{ 1
+ \frac{\alpha_s}{4 \pi} \, C_F \ln \left( \frac{Q^2}{\mu^2}
\right) \left[ -4 \sum_{j=2}^l \frac{1}{j}  - 1 + \frac{2}{l(l+1)}
\right] \Bigg\}
\nonumber \\
&+&  \frac{\alpha_s}{2 \pi} \, C_F \ln \left( \frac{Q^2}{\mu^2}
\right) \, \sum_{k=l+1}^{m} \tilde{C}_{m,k}^{(0)} \,  \left(
\frac{1}{k-l} - \frac{1}{k+1} \right),
\end{eqnarray}
for $l=1,2, \ldots, m-1$.

Let us consider a sum of products of renormalized composite
operators  and corresponding CF's which enter the OPE of two
electromagnetic currents~\eqref{02}. According to~\eqref{18}, the
renormalized composite operator $O^{m,l}_R(\mu_1^2)$ is changed
under rescaling $\mu_1 \rightarrow \mu_2$ as follows
\begin{equation}\label{206}
O^{m,l}_R(\mu_1^2)  = \sum_{l'=1}^{l} \hat{Z}^l_{l'}
(\mu_2^2/\mu_1^2) \, O^{m,l'}_R (\mu_2^2),
\end{equation}
where $\mu_i$ denotes a renormalization scale of the composite
operators in the $\overline{\mathrm{MS}}$-scheme,%
\footnote{In the renormalization scheme with subtractions,
$-\mu^2$ is off-shell renormalization point of Green functions.}
and $\mathbf{\hat{Z}}$ is a matrix of a finite renormalization.
Let us emphasize that due to Eq.~\eqref{B16}
\begin{equation}\label{207}
{\langle O^{m,l} O^{n,k} \rangle}_R (\mu_1^2) = \sum_{l'=1}^l
\hat{Z}^l_{l'} (\mu_2^2/\mu_1^2) \sum_{k'=1}^k \hat{Z}^k_{k'}
(\mu_2^2/\mu_1^2) \, {\langle O^{m,l'} O^{n,k'} \rangle}_R
(\mu_2^2),
\end{equation}
since additive coefficients in  Eq.~\eqref{B16} do not depent on
the renormalization scale $\mu$.

In the first order of strong interaction, the matrix looks like
\begin{equation}\label{208}
\hat{Z}_{l l'} =
\left\{%
\begin{array}{cl}
  \displaystyle
    1 + C_F \, \frac{\alpha_s}{4\pi} \, \ln  \left( \frac{\mu_2^2}{\mu_1^2} \right)
    \Big[ - 4 \sum\limits_{j=2}^l \frac{1}{j} - 1 + \frac{2}{l(l+1)} \Big],
    & l' = l,
  \\
  \displaystyle
    C_F \, \frac{\alpha_s}{2\pi} \, \ln  \left( \frac{\mu_2^2}{\mu_1^2} \right)
    \Big[ \frac{1}{l-l'} - \frac{1}{l+1}\Big], & l' < l
  \\
    0, & l' > l
  \\
\end{array}%
\right.
\end{equation}
In a general case we find:
\begin{eqnarray}\label{210}
\sum_{l=1}^{m} \tilde{C}_{m,l} \, O^{m,l}_R(\mu_1^2) &=&
\sum_{l=1}^{m} \tilde{C}_{m,l} \sum_{l'=1}^{l} \hat{Z}^l_{l'}
(\mu_2^2/\mu_1^2) \, O^{m,l'}_R(\mu_2^2)
\nonumber \\
&=& \sum_{l=1}^{m} \tilde{C}_{m,l}(\mu_2^2/\mu_1^2) \,
O^{m,l}_R(\mu_2^2),
\end{eqnarray}
where
\begin{equation}\label{212}
\tilde{C}_{m,l}(\mu_2^2/\mu_1^2) = \sum_{k=1}^{m} \tilde{C}_{m,k}
\, \hat{Z}^k_{l} (\mu_2^2/\mu_1^2).
\end{equation}
Thus, we obtain from \eqref{212}
\begin{equation}\label{214}
\tilde{C}_{m,m}(\mu_2^2/\mu_1^2) = \tilde{C}_{m,m} \,
\hat{Z}^m_{m} (\mu_2^2/\mu_1^2),
\end{equation}
\begin{equation}\label{216}
\tilde{C}_{m,l}(\mu_2^2/\mu_1^2) = \tilde{C}_{m,l} \,
\hat{Z}^l_{l}(\mu_2^2/\mu_1^2) + \sum_{k=l+1}^{m} \tilde{C}_{m,k}
\, \hat{Z}^k_{l}(\mu_2^2/\mu_1^2),
\end{equation}
for $l=1,2, \ldots, m-1$. It was taken into account that
$\hat{Z}_{k l} = 0$ for $k < l$. Thus, if one changes the
renormalization scale $\mu$, the ``major'' coefficient function
$\tilde{C}_{m,m}$ is simply multiplied by the factor $\hat{Z}_{m
m}$, while the other CF's mix with each other. It is a consequence
of the fact that the matrix $\hat{Z}_{kl}$ has a triangle form
with zero elements above its diagonal.

The coefficient functions $\tilde{C}_{m,l}(\mu_2^2/\mu_1^2)$ are
normalized so that $\tilde{C}_{m,l}(1) = \tilde{C}_{m,l}$, where
numbers $\tilde{C}_{m,l}$ are ``bare'' coefficient functions which
were obtained for the case with no strong interactions (see
Eqs.~\eqref{107}, \eqref{108}). We conclude from explicit
expression of $\mathbf{\hat{Z}}$ in the first order in $\alpha_s$
\eqref{208} that our perturbative result, equations \eqref{202}
and \eqref{204}, is a particular
case of general formulae \eqref{214} and \eqref{216}.%
\footnote{After an obvious replacement $\mu_2^2/\mu_1^2
\rightarrow Q^2/\mu^2$.}


\section{Conclusions}

As was shown above, the OPE coefficient functions can be expressed
in terms of the Green functions of the corresponding composite
operators without explicit use of the elementary (quark and gluon)
field operators. We would like to specially stress that our
general expression \eqref{56} for the OPE coefficient functions
holds in \emph{any} renormalization scheme in contrast with
previous prescriptions (see, for instance, \cite{Chetyrkin:82}).
We believe that such a representation is useful in proving general
properties of the coefficient functions and for non-perturbative
calculations (e.g. in lattice QCD).

We would like to stress that if one assumes the light-cone
expansion in the framework of axiomatic approach (see, for
instance, Ref.~\cite{Wilson:72}), then our formulae do not lean on
perturbative expansions at all and deal with v.e.v. of T-products
of local Heisenberg operators.

In this paper we limited ourselves by the QCD nonsinglet operator
only. The singlet case needs a special treatment and will be
considered in a forthcoming paper.



\setcounter{equation}{0}
\renewcommand{\theequation}{A.\arabic{equation}}

\section*{Appendix A}
\label{app:A}

In this Appendix we have collected formulae which are needed for
calculating sums in $l$ and $j$ presented in the text and getting
compact expressions. Let us first consider summation in index $l$.
For $m \geqslant 0$, we have the relation
\begin{equation}\label{A01}
\sum_{l=0}^{m} (-1)^l {m\choose{l}} {l+k+1\choose{l+1}}^{-1}
\frac{1}{l+1} = \frac{1}{m+k+1} \, .
\end{equation}
The quantity ${n\choose{k}}$ is a binomial coefficient.%
\footnote{The beta-function is given by the equation
$[B(n+1,k+1)]^{-1} = (n+k+1)\,{n+k\choose{k}}$.}
In the following equations, $m \geqslant 1$ is assumed:
\begin{equation}\label{A02}
\sum_{l=1}^{m-1} (-1)^l {m-1\choose{l-1}} = - \left[(-1)^m +
\frac{1}{\Gamma(2-m)} \right],
\end{equation}

\begin{equation}\label{A04}
\sum_{l=1}^{m-1} (-1)^l {m-1\choose{l-1}} \! \frac{1}{l} = -
\frac{1}{m}[1 + (-1)^m],
\end{equation}

\begin{equation}\label{A06}
\sum_{l=1}^{m-1} (-1)^l {m-1\choose{l-1}} \! \frac{1}{l+1} = -
\frac{1}{m} + \frac{1}{m+1} [1 - (-1)^m],
\end{equation}

\begin{equation}\label{A08}
\sum_{l=1}^{m-1} (-1)^l {m-1\choose{l-1}} \! \frac{1}{(l+1)(l+2)}
= - \frac{1}{(m+1)(m+2)}[1 + (-1)^m],
\end{equation}

\begin{eqnarray}\label{A10}
&& -2 \, \sum_{l=1}^{m-1} \left( \frac{1}{m-l} - \frac{1}{m+1}
\right) \frac{1}{(l+1)(l+2)}
\nonumber \\
&& = \frac{1}{(m+1)(m+2)} \left[ -4 \sum_{j=2}^{m} \frac{1}{j} -1
+ \frac{2}{m(m+1)} \right],
\end{eqnarray}

\begin{eqnarray}\label{A12}
&& \sum_{l=1}^{m-1} (-1)^l {m-1\choose{l-1}}
{l+k+1\choose{l+1}}^{-1}
\nonumber \\
&& = (-1)^{m+1} {m+k+1\choose{m+1}}^{-1} \! \frac{1}{m+1} -
\frac{1}{(m+k)(m+k+1)} \, ,
\end{eqnarray}

\begin{eqnarray}\label{A14}
&& \sum_{l=1}^{m-1} (-1)^l {m-1\choose{l-1}}
{l+k+1\choose{l+1}}^{-1}  \frac{1}{l}
\nonumber \\
&& = (-1)^{m+1} {m+k+1\choose{m+1}}^{-1}  \frac{1}{m} +
\frac{k}{(k+1)(m+k+1)} - \frac{1}{m+k} \, ,
\end{eqnarray}

\begin{eqnarray}\label{A16}
&& \sum_{l=1}^{m-1} (-1)^l {m-1\choose{l-1}}
{l+k+1\choose{l+1}}^{-1} \! \frac{1}{l(l+1)}
\nonumber \\
&& = (-1)^{m+1} {m+k+1\choose{m+1}}^{-1} \! \frac{1}{m(m+1)} -
\frac{1}{(k+1)(m+k+1)} \, .
\end{eqnarray}

Now let us consider summation in index $j$ ($k,l \geqslant 1$ is
assumed everywhere):
\begin{equation}\label{A20}
\sum_{j=1}^{k} {m+k-j-1\choose{m-1}}  \frac{1}{j} =
{m+k-1\choose{k}} \left[ \sum_{j=1}^{m+k-1} \frac{1}{j} -
\sum_{j=1}^{m-1} \frac{1}{j} \right],
\end{equation}

\begin{equation}\label{A22}
\sum_{j=1}^{m-1} \frac{1}{j} - \sum_{j=1}^{l-1} \frac{1}{j} =
{m-1\choose{l-1}}^{-1} \sum_{j=1}^{m-l} {m-j-1\choose{l-1}}
\frac{1}{j} \, .
\end{equation}
Note that the last equation is a particular case of
Eq.~\eqref{A20} for $1 \leq l \leq m$. Let us present other useful
formulae:
\begin{equation}\label{A24}
\sum_{j=1}^{l-1} {k+j+1\choose{j}}^{-1} = - \frac{k+l+1}{k}
{k+l+1\choose{l}}^{-1} + \frac{1}{k} \, ,
\end{equation}

\begin{eqnarray}\label{A26}
\sum_{j=0}^{l-1} {k+j+1\choose{j}}^{-1} \! \frac{1}{l-j} &=&
{k+l+1\choose{l}}^{-1}
\nonumber \\
&\times& \left[ \sum_{j=1}^{l} \frac{1}{j} + \sum_{j=1}^{l}
{k+j\choose{j}} \, \frac{1}{j} \right],
\end{eqnarray}

\begin{equation}\label{A28}
\sum_{j=1}^{l-1} \frac{1}{j} \left[ {k+j\choose{j}} - 1 \right] =
\sum_{j=1}^{k-1} \frac{1}{j} \left[ {l+j\choose{j}} - 1 \right],
\end{equation}

\begin{eqnarray}\label{A30}
&& \frac{1}{k+1} \left[ \sum_{j=0}^{l-1} {k+j+1\choose{j}}^{-1} \!
\frac{1}{l-j} - 2 \, {k+l+1\choose{l}}^{-1} \sum_{j=2}^{l}
\frac{1}{j} \right]
\nonumber \\
&& = \! \frac{1}{l+1} \left[ \sum_{j=0}^{k-1}
{l+j+1\choose{j}}^{-1} \! \frac{1}{k-j} - 2 \,
{l+k+1\choose{k}}^{-1} \sum_{j=2}^{k} \frac{1}{j}  \right].
\end{eqnarray}
The last relation is a consequence of Eqs.~\eqref{A26} and
\eqref{A28}. It demonstrates that the sum in \eqref{A30} is
symmetric under replacement $l \rightleftarrows k$.

The last useful formula contains sums in $l$ and $j$:
\begin{eqnarray}\label{A32}
&& \sum_{l=1}^{m-1} (-1)^l {m-1\choose{l-1}}
{l+k+1\choose{l+1}}^{-1} \! \frac{1}{l+1} \, \sum_{j=1}^{k}
{l+j\choose{j}} \frac{1}{j}
\nonumber \\
&& = - \, {m+k+1\choose{m+1}}^{-1} \! \frac{1}{m+1} \left[
\sum_{j=0}^{k-1} {m+j-1\choose{j}} \frac{k-j+1}{(k-j)(m+j)}
\right.
\nonumber \\
&& + \, (-1)^m \left.   \sum_{j=1}^{k} {m+j\choose{j}} \frac{1}{j}
\right].
\end{eqnarray}

\clearpage

\setcounter{equation}{0}
\renewcommand{\theequation}{B.\arabic{equation}}

\section*{Appendix B}
\label{app:B}

The goal of this Appendix is to derive a renormalization group
equation (see, for instance, \cite{Bogolyubov}) for the CF's
stating from our formula~\eqref{56}. Let us
rewrite it in the form:%
\footnote{Here and in what follows the subscript $_{R(U)}$ means
that a corresponding quantity is a renormalized (unrenormalized)
one.}
\begin{equation}\label{B02}
\sum_{l=1}^m \tilde{C}_{m,l} \, \langle O O \rangle^{kl}_R =
\frac{1}{2\pi i} \int\limits_0^{1} \! d x x^{m-1} \, \mbox{\rm
disc}_{(p+q)^2} \, \langle J J O \rangle^{k}_R
\end{equation}
(see our brief notations in the end of Section~\ref{sec:OPE}). The
renormalized quantity in the RHS of Eq.~\eqref{B02} is given by
\begin{equation}\label{B04}
\sum_n \langle J |n \rangle \, \langle n | J O^{n,k} \rangle_R,
\end{equation}
where $\sum_n$ means a sum in a complete set of elementary (quark
and gluon) states $|n \rangle$. Let us underline that each state
$|n \rangle$ should include at least one quark-antiquark pair (in
non-singlet state). It is well-known that a matrix element with an
insertion of \emph{one} composite operator is multiplicatively
renormalized (see, for instance, \cite{Itzykson}). Therefore, one
has the following relation:
\begin{equation}\label{B06}
\langle J |n \rangle_R = Z_J \langle J |n \rangle_U.
\end{equation}
Since electromagnetic current is conserved, its renormalization
constant, $Z_J$, does not depend on the renormalization scale,
$(d/d\mu)Z_J = 0$. ,

Now we will show that the matrix element $\langle n | J O^{n,k}
\rangle_R$ in \eqref{B04} is also multiplicatively renormalized.
Consider arbitrary diagram $G$ which contributes to a matrix
element with an insertion of two composite operators, $J$ and
$O^k$. We assume that divergences of all sub-diagrams of $G$ are
already removed. Let $G$ to have $(p+2)$ external lines ($p
\geqslant 0$), and $q$ internal vertexes. The index of this
diagram, $\omega (G)$, is defined by~\cite{Itzykson}
\begin{equation}\label{B08}
\omega (G) = \sum_i^q \omega_i^{max} + 4 - \frac{1}{2}
\sum_{l}^{p+2} (r_l + 2) + (\omega_J - 4) + (\omega_{O^{n,k}} -
4).
\end{equation}
Here $\omega_i^{max}$ is a maximal index of internal vertex of a
type $i$, and $r_l$ is a power of a polynomial corresponding to an
external field of a type $l$ ($r_q=1$ for a quark line). The
quantities $\omega_J$ is the dimension of the electromagnetic
current, while $\omega_{O^{n,k}}$ is the dimensions of our
composite operators. Note that $\omega_i^{max} = 0$ for all QCD
vertexes, and $\omega_J = 3$.

As for $\omega_{O^{n,k}}$, it is \emph{formally} equal to $(k+2)$
(only derivatives with respect to quark line momenta have to be
taken into account, not total derivatives). In particular,
$\omega_{O^{n,1}} = 3$. Let us show, however, that effectively
$\omega_{O^{n,k}}$ does not depend on $k$, and $\omega_{O^{n,k}} =
\omega_{O^{n,1}}$. Let $\{r_i \}$ ($i = 1, \ldots p+4$) to be a
set of external momenta ($p, q$ including), while $\{ k_i \}$ ($i
= 1, \ldots m$) to be a set of loop momenta of $G$ (with $m$ being
a number of loops). After using the Feynman parametrization and
\emph{linear} transformations from $\{ k_i \}$ to $\{ l_i \}$, an
analytical expression for the diagram can be written in the form:
\begin{eqnarray}\label{B09}
I(G) &=&  \prod_j \int \! d x_j \, \delta (1 - \sum_s x_s )
\sum_{r=0}^{k-1} \, a_r (x_i)
\nonumber \\
&\times& \prod_{i=1}^m \int \! \frac{d^D l_i}{(2\pi)^D} \, (l_m
n)^{r} \, \frac{P (x_i, l_i, r_i)}{D (x_i, l_i^2, r_ir_j)} \ ,
\end{eqnarray}
where $x_i$ are the Feynman parameters. The change of variables
$\{ k_i \} \rightarrow \{ l_i \}$ is done so that the denominator
in Eq.~\eqref{B09} depends on $l_i$ only via $l_i^2$. Notice, the
momentum $l_m$ is a linear combination of $r_i$, and $k_m$,
ingoing quark momentum for the vertex $O^{n,k}$. The numerator in
Eq.~\eqref{B09} is a sum of polynomials of $(l_il_j)$, $(l_ir_j)$
and $(r_ir_j)$. Each polynomial includes at least one scalar
product of the vector $n$, namely, $(l_in)$ or $(r_in)$.

After integrating in $d^Dl_1, \dots d^Dl_{m-1}$, we get
\begin{eqnarray}\label{B10}
I(G) &=& \prod_j \int \! d x_j \, \delta (1 - \sum_s x_s )
\sum_{r=0}^{k-1} \, b_r (x_i)
\nonumber \\
&\times& \int \! \frac{d^D l}{(2\pi)^D} \, (ln)^{r}\,
\frac{\bar{P} (x_i, l, r_i)}{\tilde{D}(x_i, l^2, r_ir_j)}
\end{eqnarray}
(we denoted $l_m = l$). Note that $\mathrm{dim}\bar{D} -
\mathrm{dim}\bar{P} = 5$. Let us analyze possible terms in
$\bar{P}$  which is a polynomial of $l^2$, $(lr_i)$, $(r_ir_j)$,
and depends linearly on $(ln)$ or $(r_in)$. The term
$(ln)(l^2)^{(\mathrm{dim}\bar{P}-1)/2}$ results in zero (for even
$r$) or an expression proportional to $n^2_{\mu}
=0$ (for odd $r$) in the dimensional regularization.%
\footnote{Similar arguments are used under calculations of the
anomalous dimensions of the composite operators (see, for
instance, Ref.~\cite{Buras:82}).}
All other terms proportional to $(ln)$ drop for the same reason
(if $r \geqslant 1$) or give finite integrals which converge in
the ultra-violet region. One can verify that terms proportional to
$(r_in)$ give non-zero convergent integrals only for $r=0$.

Thus, we  have to put $\omega_{O^{n,k}} = \omega_{O^{n,1}} = 3$ in
\eqref{B08}. In its turn, it means that $\omega (G) \leqslant -1$,
with $\omega (G) = -1$ for $p=1$. In other words, there is no need
to introduce a new counterterm $:J(x)O(y): \delta (x-y)$, and,
consequently,
\begin{equation}\label{B12}
\langle n | J O^{n,k} \rangle_R = Z_J^{-1} \sum_{k'=1}^k
(Z^{-1})^k_{k'} \, \langle n | J O^{n,k'} \rangle_U.
\end{equation}
It follows from \eqref{B06}, \eqref{B12} that
\begin{equation}\label{B14}
\mathrm{disc}_{(p+q)^2} \, \langle J J O \rangle^{k}_R = Z_J^{-2}
\sum_{k'=1}^k (Z^{-1})^k_{k'} \, \mathrm{disc}_{(p+q)^2} \,
\langle J J O \rangle^{k'}_U.
\end{equation}
This result is in agreement with perturbative QCD calculations
from Section~\ref{sec:CF_QCD}.

Now let us turn to the propagators of the composite operators. The
renormalization properties of Green functions with insertions of
more than one composite operator were studied in details in
Ref.~\cite{Shore:91}. In particular,
\begin{equation}\label{B16}
\langle O O \rangle^{kl}_R = \sum_{l'=1}^l (Z^{-1})^l_{l'}
\sum_{k'=1}^k (Z^{-1})^k_{k'} \Big[ \langle O O \rangle^{k'l'}_U +
f_B^{k'l'} \Big].
\end{equation}
The divergent coefficients $f_B^{kl}$ are, in general, non-zero
even in the free theory, as it is the case for our quark operators
$Q^{n,k}$.

The renormalization matrix of the composite operator $\mathbf{Z}$
depends on a \emph{renormalization} scale $\mu$, while the
coefficients $f_B^{kl}$ do not. The renormalized Green function
$\langle O O \rangle^{kl}_R$ is also $\mu$-dependent, but it does
not depend on the \emph{regularization} scale $\bar{\mu}$. All
physical (measurable) quantities should be, of course, $\bar{\mu}$
and $\mu$-independent.

As one can see, Eq.~\eqref{B16} which relates renormalized and
unrenormalized  propagators has an additive term. Nevertheless,
renormalization group equations for the propagators and for
CF's have no additive terms.%
\footnote{However, the renormalization group equations are not
homogeneous if some of the composite operators has non-zero
vev~\cite{Shore:91}. In our case, $\langle O^{n,k} \rangle = 0$,
and $O^{n,k}$ do not mix with the identity operator under the
renormalization.}
Indeed, acting on both sides of Eq.~\eqref{B02} by the operator
$\mu d/d\mu = \mu
\partial / \partial \mu + \beta (g_s) \partial /
\partial g_s$, and taking into account Eqs.~\eqref{B14}, \eqref{B16},
we obtain:
\begin{equation}\label{B18}
\sum_{l=1}^m \Big[ \mu \frac{d}{d\mu} \, \tilde{C}_{m,l} \,
\langle O O \rangle^{kl}_R - \tilde{C}_{m,l}  \sum_{l'=1}^l
\gamma^l_{l'} \, \langle O O \rangle^{kl'}_R  \Big] = 0,
\end{equation}
where
\begin{equation}\label{B20}
\boldsymbol{\gamma} = \mathbf{Z}^{-1} \, \Big( \mu'
\frac{d}{d\mu'} \Big) \mathbf{Z}
\end{equation}
is the matrix of anomalous dimensions of the composite operators.
Equation~\eqref{B18} can be represented in the form:
\begin{equation}\label{B22}
\sum_{l=l}^m \langle O O \rangle^{kl}_R \, \Big[ \mu'
\frac{d}{d\mu'} \, \tilde{C}_{m,l} - \sum_{l'=l}^m \gamma_l^{l'}
\, \tilde{C}_{m,l'} \Big] = 0.
\end{equation}
Since \eqref{B22} is valid for \emph{arbitrary} integer $k
\geqslant 1$, we derive a set of renormalization group equations
for the coefficient functions $\tilde{C}_{m,l}$ ($l=1,2, \ldots
m$):
\begin{equation}\label{B24}
\mu \frac{d}{d\mu} \, \tilde{C}_{m,l} - \sum_{l'=l}^m
\gamma_l^{l'} \, \tilde{C}_{m,l'}  = 0.
\end{equation}
In particular, the coefficient function $\tilde{C}_{m,m}$, which
is relevant for DIS, obeys the following closed equation:
\begin{equation}\label{B26}
\Big( \mu \frac{d}{d\mu}  - \gamma_m^m \Big) \tilde{C}_{m,m}  = 0.
\end{equation}

Remember that $\boldsymbol{\gamma}$ is the matrix of the anomalous
dimensions of non-singlet composite operators which enter
light-cone OPE~\eqref{10}:
\begin{equation}\label{B28}
\mu \frac{d}{d\mu} \, O^{m,l} + \sum_{l'=1}^l \gamma^l_{l'} \,
O^{m,l'}  = 0.
\end{equation}
This renormalization group equation demonstrates again that
composite operators mix under the renormalization. After a
diagonalization of the renormalization matrix
$\mathbf{Z}$~\eqref{20}, one obtains new operators
$\tilde{O}^{m,l} = \sum_{l'=1}^l A_{l'}^l \, O^{m,l'}$ which are
multiplicatively renormalized. The matrix $\mathbf{A}$ has a
triangle form, with $\mathrm{det} \mathbf{A} \neq 0$. The
anomalous dimensions of the operators $\tilde{O}^{m,l}$ are
determined by diagonal elements of $\mathbf{Z}$. The one-loop
values of $\tilde{\gamma}_l = \gamma_l$ are given by
Eq.~\eqref{142}.

The analysis of renormalization properties of the composite
operator propagator is simplified, if we choose a light-cone axial
gauge $n^{\mu} A_{\mu} = 0$ in which only diagrams shown in
Fig.~\ref{fig:loop_renorm} contribute to $\langle O O
\rangle^{kl}$. Correspondingly, the renormalization of the
composite operator in this gauge is given by diagram in
Fig.~\ref{fig:operator_renorm}. The blob in these figures is a sum
of all possible QCD diagrams (with a disconnected part included).
A specific character of the diagrams in
Fig.~\ref{fig:loop_renorm}, \ref{fig:operator_renorm} is the
following: they can be divided in two parts by cutting \emph{two
quark lines}. It enables us to derive the following relation:
\begin{equation}\label{B30}
\langle O O \rangle_U^{kl}  = A^{kl}_{k'l'} \left[ \varepsilon,
\left ( \bar{\mu}^2/p^2 \right)^{\varepsilon} \right] \,
Z_B^{k'l'} (\varepsilon, \bar{\mu}^2/p^2),
\end{equation}
where the matrix $\mathbf{Z_B}(\varepsilon, \bar{\mu}^2/p^2)$ is a
bare quark loop with an insertions of two composite operators:
\begin{equation}\label{B32}
Z_B^{kl} (\varepsilon, \bar{\mu}^2/p^2) \sim \frac{1}{\varepsilon}
\left( \frac{\bar{\mu}^2}{-p^2} \right)^{\varepsilon} B(k+1,l+1)
\end{equation}
(terms which are non-singular at $p^2 \rightarrow 0$ are omitted
in \eqref{B32}). Thus, the coefficients $f_B^{kl}$ in \eqref{B16},
which are needed to regularized the Green functions of two
composite operators, look like
\begin{equation}\label{B34}
f_B^{kl} \sim - \frac{1}{\varepsilon} \, A^{kl}_{k'l'}
(\varepsilon, 1) \, B(k'+1,l'+1).
\end{equation}
The relation~\eqref{B30} is confirmed by  our calculations in
perturbation theory (see, for instance, one-loop \eqref{102} and
two-expressions \eqref{115} for $\langle O O \rangle_R$).


\clearpage


\clearpage

\clearpage

\clearpage

\newcounter{subfigure}

\renewcommand{\thesubfigure}{\alph{subfigure}}

{
\renewcommand{\thefigure}{\arabic{figure}\alph{subfigure}}

\setcounter{subfigure}{1}

\begin{figure}[ht]
\resizebox{\textwidth}{!}{\includegraphics{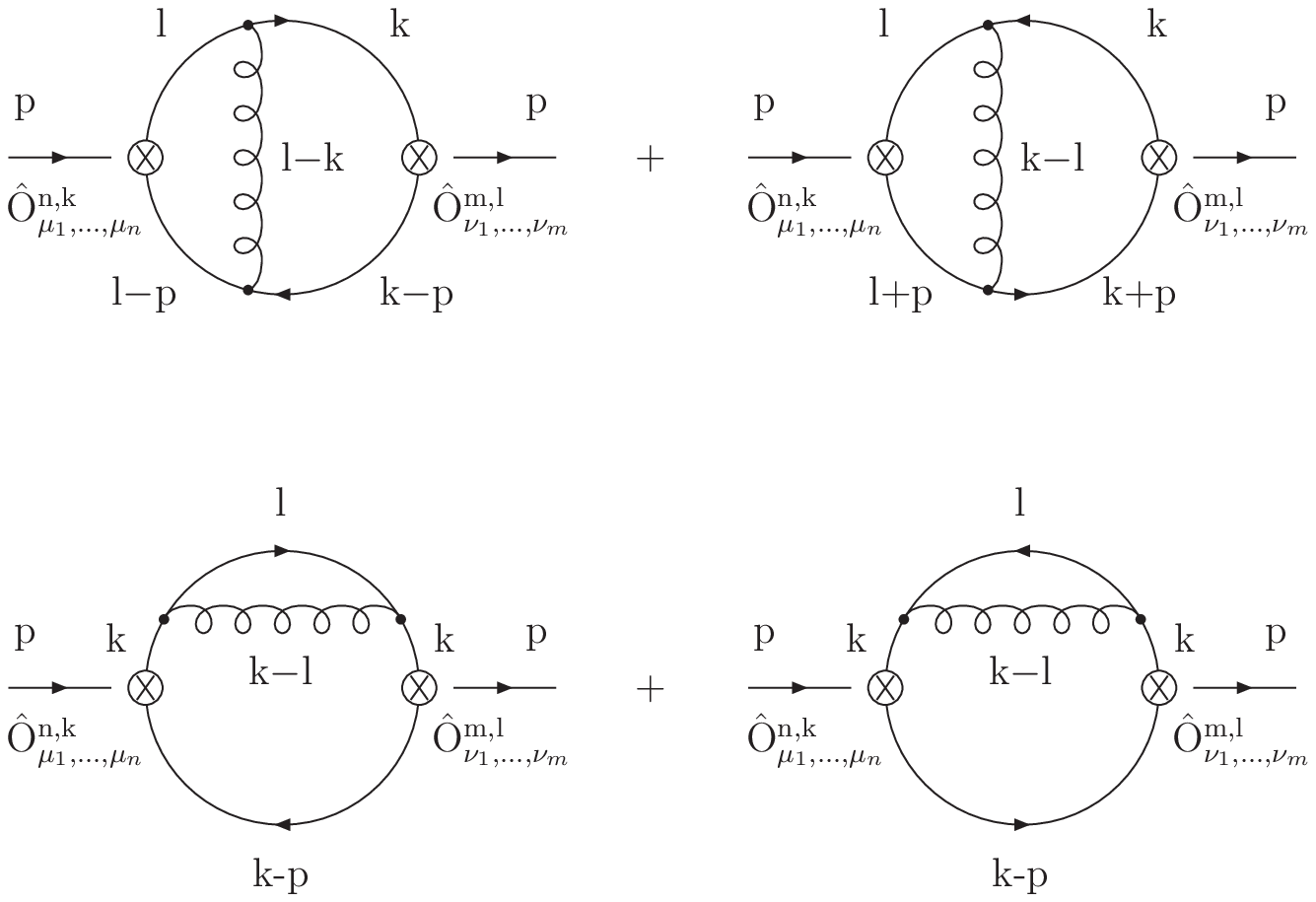}} \caption{The
diagrams for the propagator $\langle O^{n,k} O^{m,l}
\rangle^{(1)}$.} \label{fig:loop_1_a}
\end{figure}

\addtocounter{figure}{-1} \setcounter{subfigure}{2}

\begin{figure}[ht]
\resizebox{\textwidth}{!}{\includegraphics{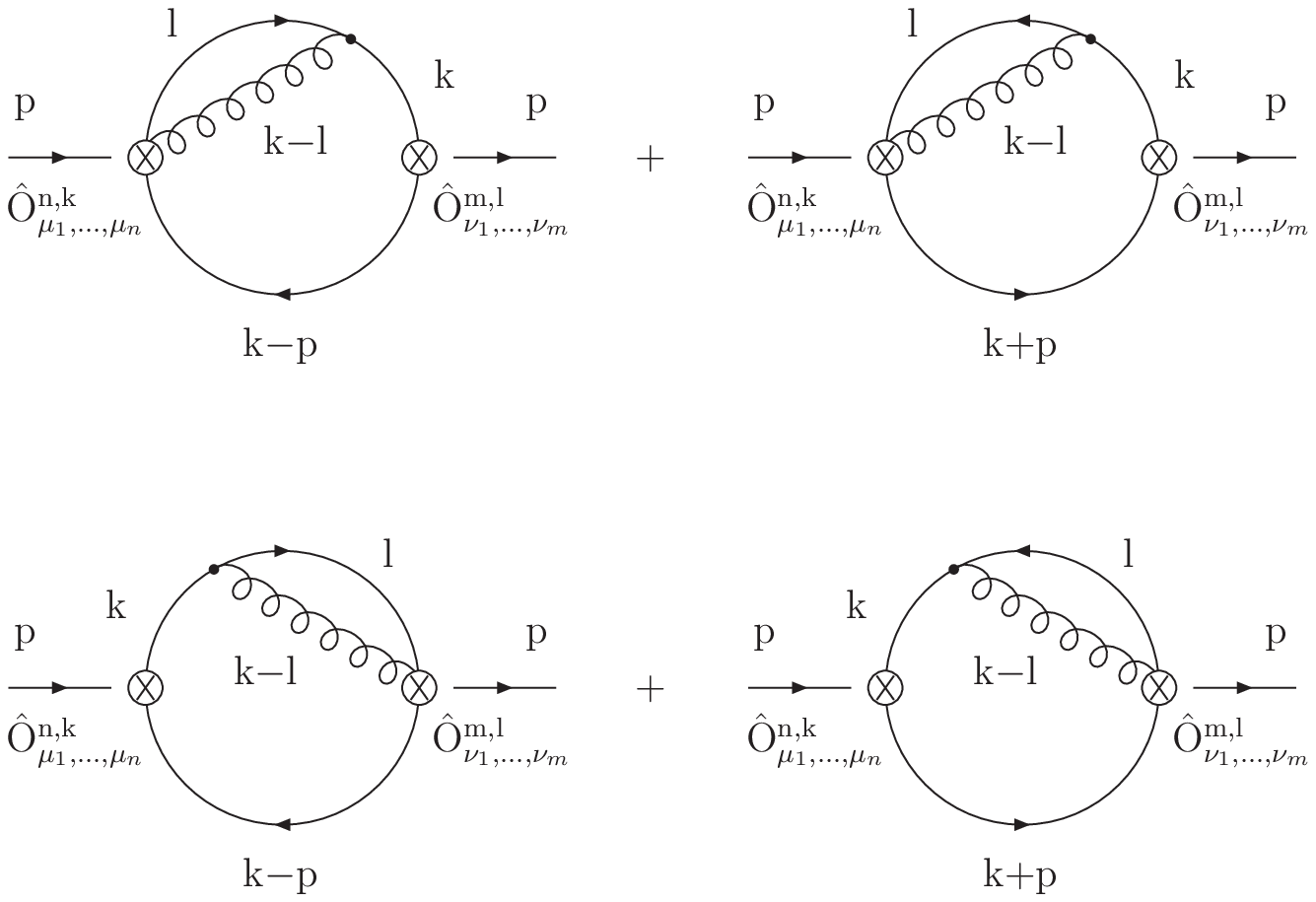}}
\caption{The diagrams for the propagator  $\langle O^{n,k} O^{m,l}
\rangle^{(1)}$ (continued).} \label{fig:loop_1_b}
\end{figure}
}

\begin{figure}[ht]
\resizebox{\textwidth}{!}{\includegraphics{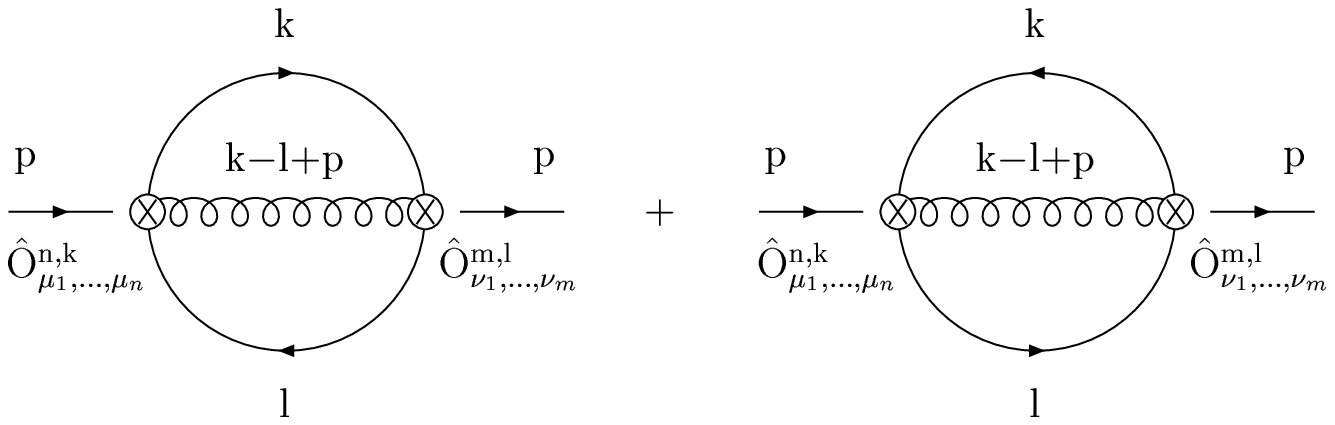}} \caption{The
diagrams which give zero contribution to the propagator $\langle
O^{n,k} O^{m,l} \rangle^{(1)}$.} \label{fig:loop_1_c}
\end{figure}

\clearpage
{
\renewcommand{\thefigure}{\arabic{figure}\alph{subfigure}}

\setcounter{subfigure}{1}

\begin{figure}[ht]
\resizebox{.9\textwidth}{!}{\includegraphics{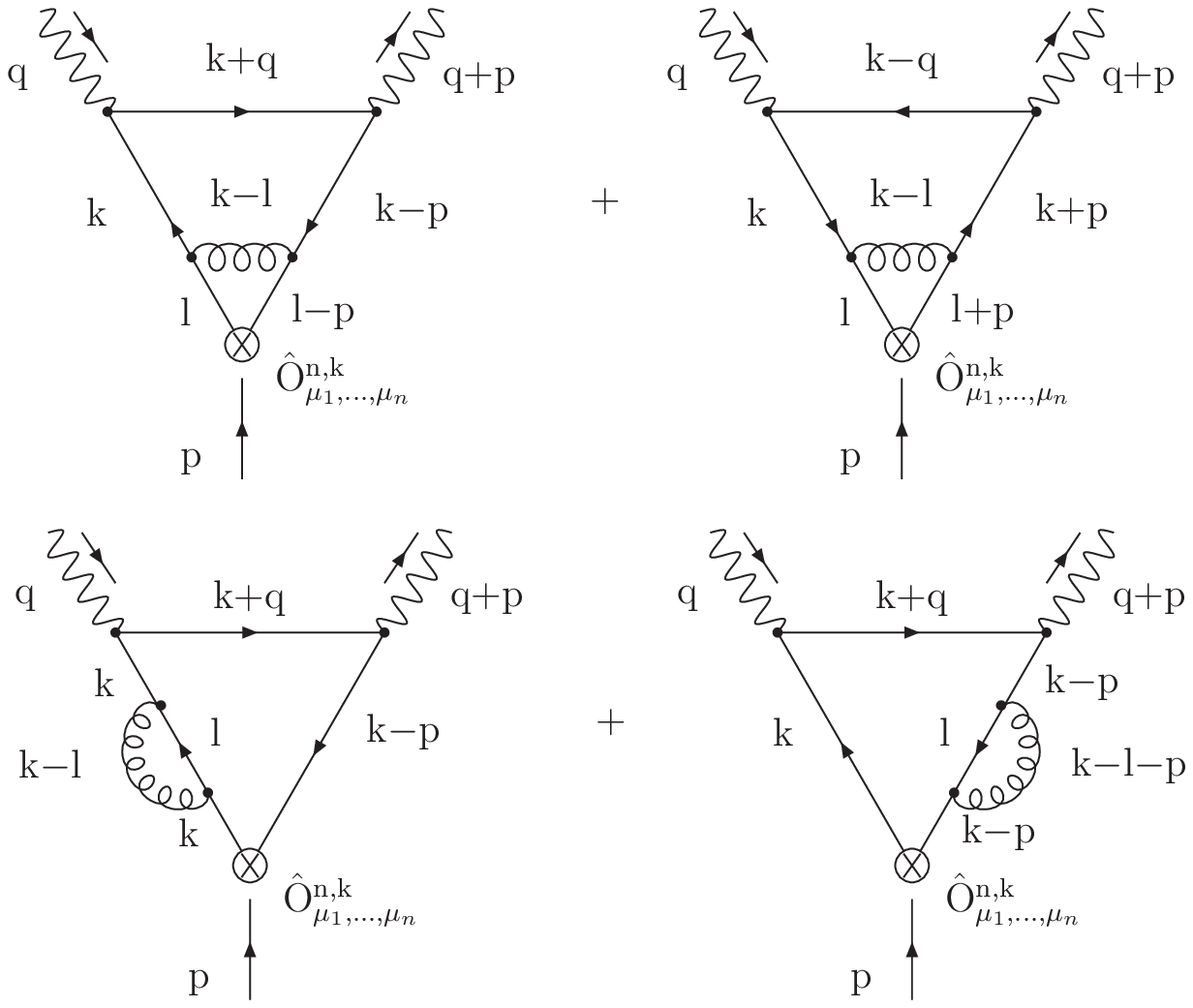}} \caption{The
diagrams for the matrix element $\langle J J O^{n,k}
\rangle^{(1)}$.} \label{fig:vertex_1_a}
\end{figure}

\addtocounter{figure}{-1} \setcounter{subfigure}{2}

\begin{figure}[ht]
\resizebox{.9\textwidth}{!}{\includegraphics{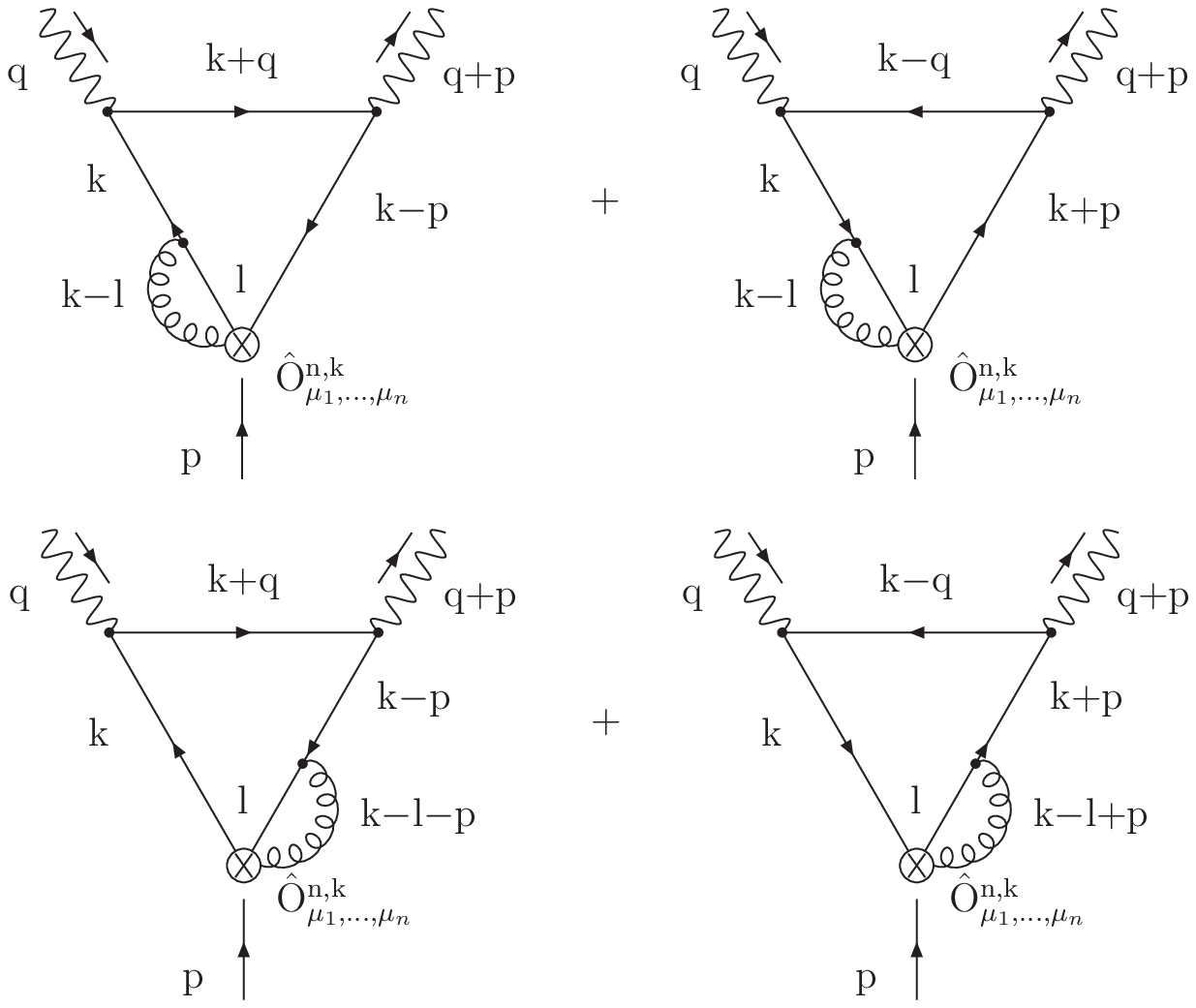}} \caption{The
diagrams for the matrix element $\langle J J O^{n,k}
\rangle^{(1)}$ (continued).} \label{fig:vertex_1_b}
\end{figure}
}

\begin{figure}[ht]
\resizebox{.9\textwidth}{!}{\includegraphics{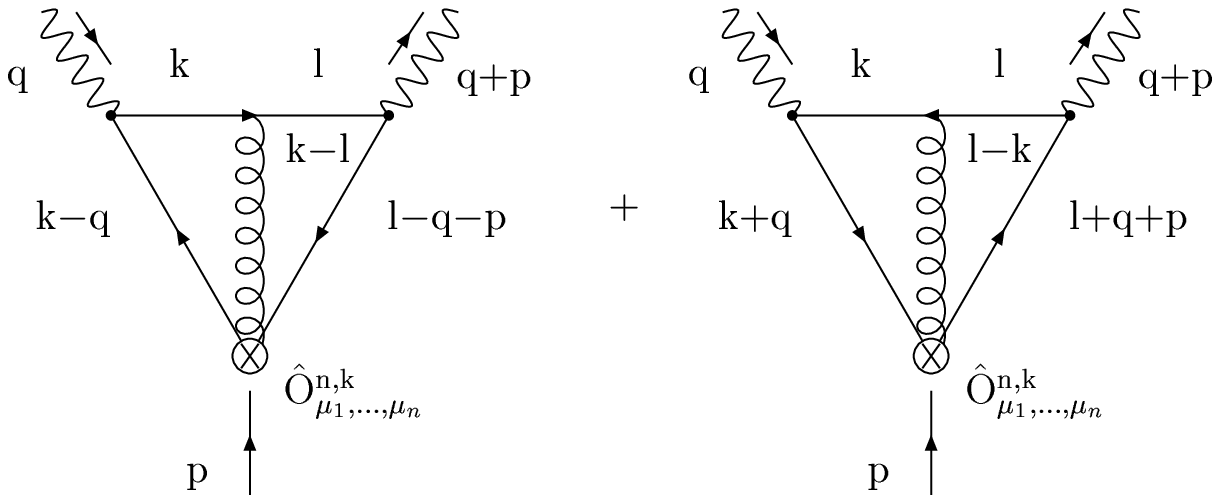}} \caption{The
diagrams which give non-leading contribution to the matrix element
$\langle J J O^{n,k} \rangle^{(1)}$ at $p^2 \rightarrow 0$.}
\label{fig:vertex_1_c}
\end{figure}

\clearpage
{
\renewcommand{\thefigure}{\arabic{figure}\alph{subfigure}}

\setcounter{subfigure}{1}

\begin{figure}[ht]
\resizebox{.9\textwidth}{!}{\includegraphics{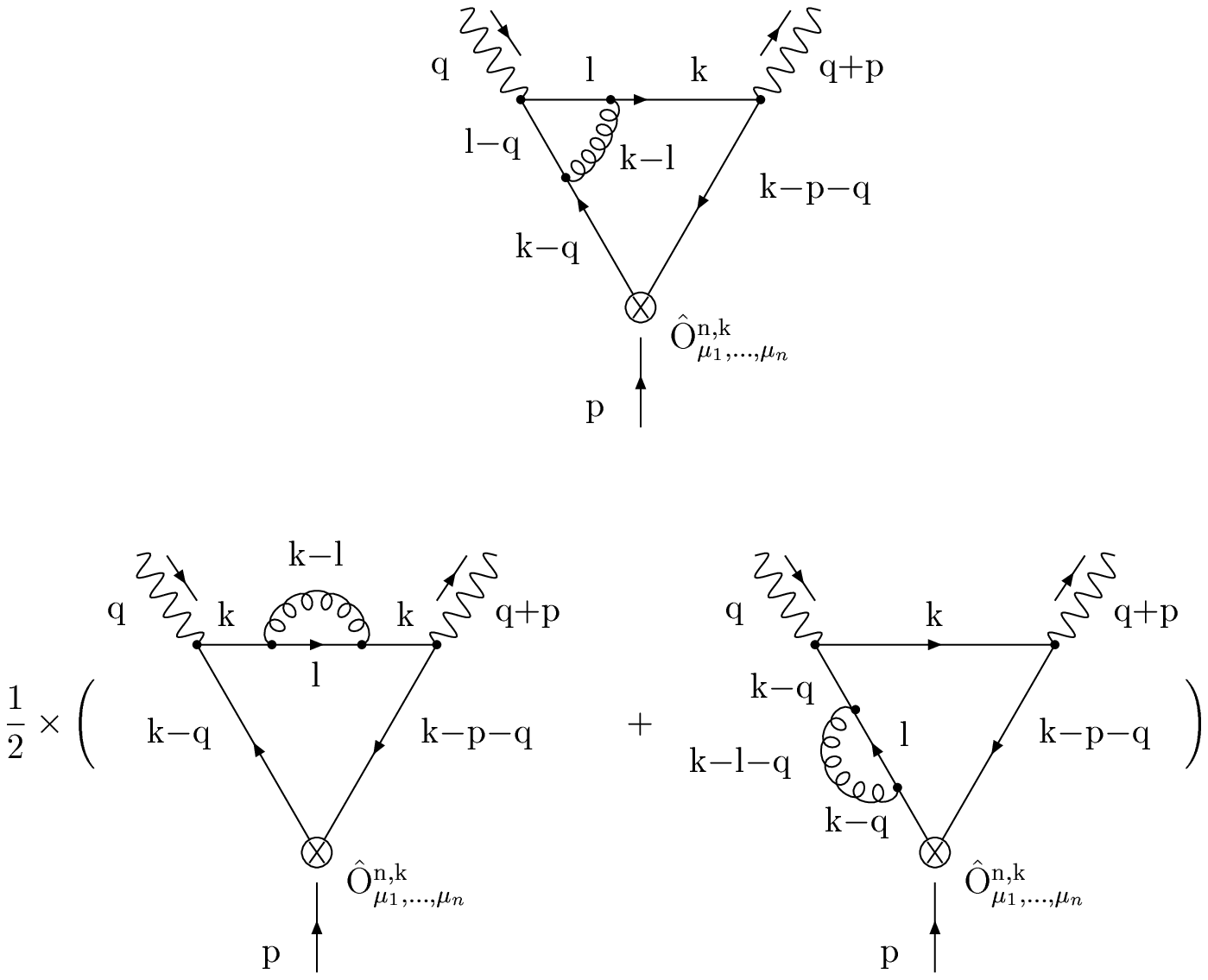}} \caption{The
diagrams which give contribution to the renormalization of the
electromagnetic current $J^{em}$.} \label{fig:current_1_a}
\end{figure}

\addtocounter{figure}{-1} \setcounter{subfigure}{2}

\begin{figure}[ht]
\resizebox{.9\textwidth}{!}{\includegraphics{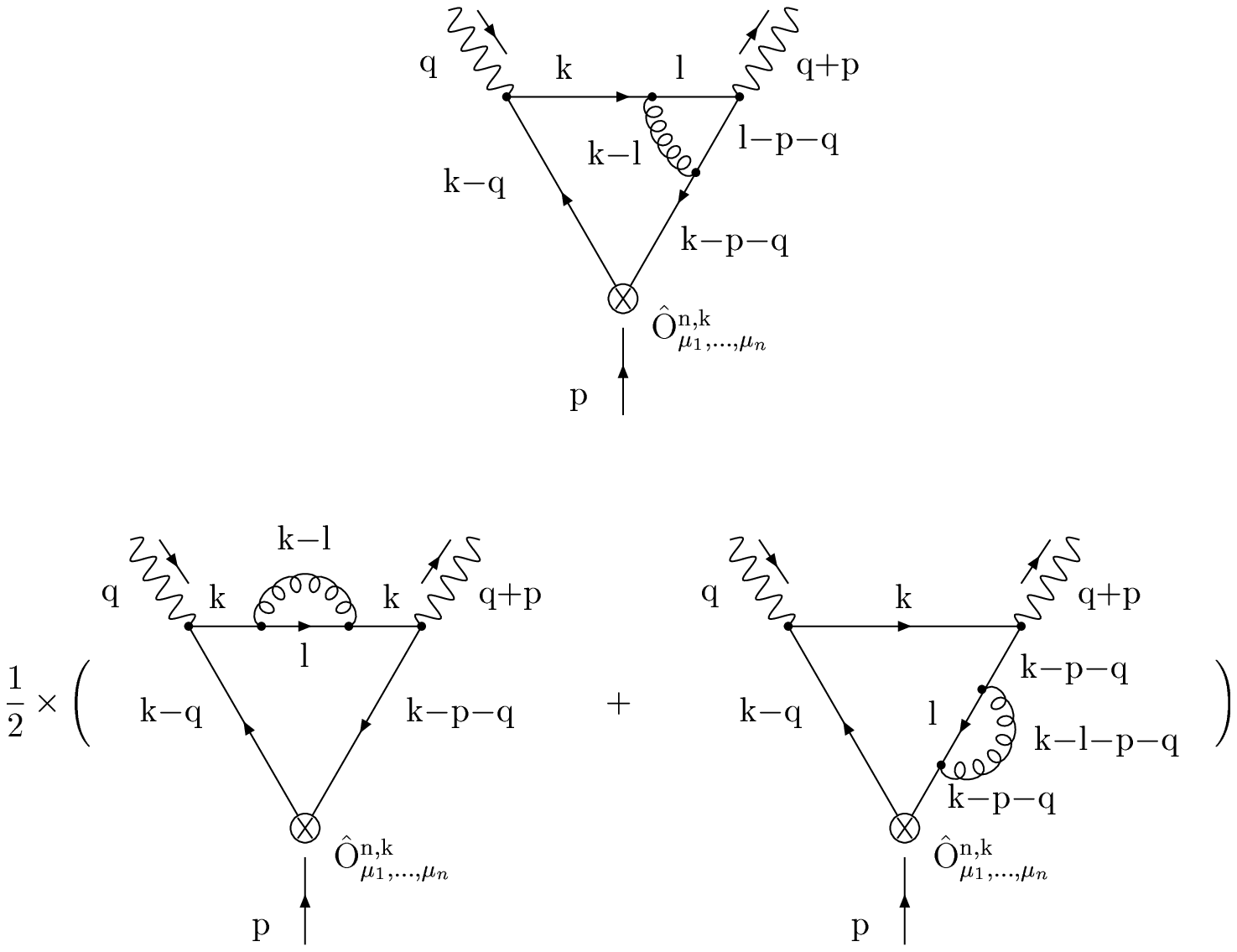}} \caption{The
diagrams which give contribution to the renormalization of the
electromagnetic current $J^{em}$ (continued).}
\label{fig:current_1_b}
\end{figure}
}

\clearpage

\begin{figure}[ht]
\resizebox{.9\textwidth}{!}{\includegraphics{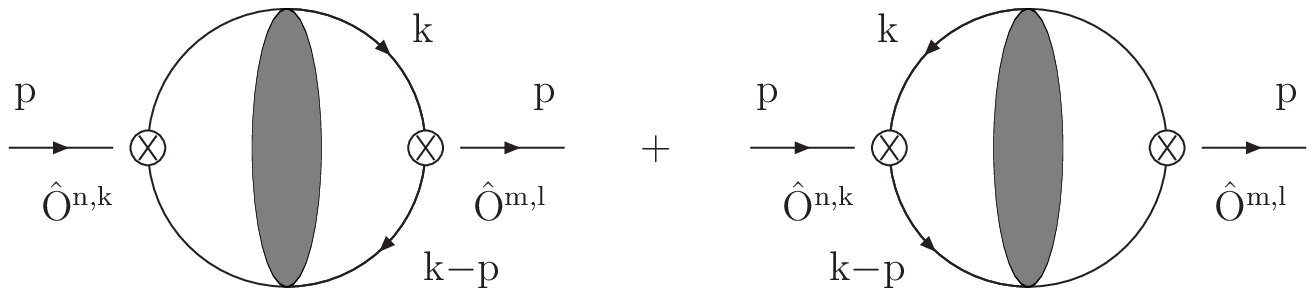}} \caption{The
diagrams which give contribution to the propagator of the
composite operator in the axial gauge.} \label{fig:loop_renorm}
\end{figure}


\begin{figure}[ht]
\hspace*{4.5cm}
\resizebox{.25\textwidth}{!}{\includegraphics{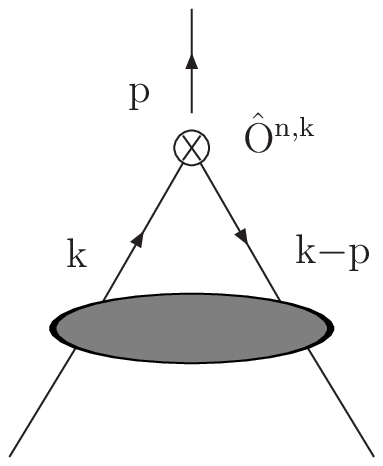}}
\caption{The diagram which describes the renormalization of the
composite operator in the axial gauge.}
\label{fig:operator_renorm}
\end{figure}

\end{document}